\documentclass[sigconf]{acmart}
\AtBeginDocument{%
  }

\copyrightyear{2026}
\acmYear{2026}
\setcopyright{cc}
\setcctype{by}
\acmConference[CHI '26]{Proceedings of the 2026 CHI Conference on Human Factors in Computing Systems}{April 13--17, 2026}{Barcelona, Spain}
\acmBooktitle{Proceedings of the 2026 CHI Conference on Human Factors in Computing Systems (CHI '26), April 13--17, 2026, Barcelona, Spain}
\acmPrice{}
\acmDOI{10.1145/3772318.3790470}
\acmISBN{979-8-4007-2278-3/2026/04}

\acmSubmissionID{5217}

\usepackage{subcaption}
\usepackage{graphicx}
\usepackage{multirow}
\usepackage{xcolor}

\newif\ifshowchanges
\showchangesfalse

\ifshowchanges
  \usepackage[markup=custom,commandnameprefix=ifneeded]{changes}

  \definecolor{mygreen}{RGB}{0, 120, 0}
  \definecolor{myred}{RGB}{180, 50, 50}

  \setaddedmarkup{\textcolor{mygreen}{#1}}
  \setdeletedmarkup{\textcolor{myred}{\sout{#1}}}

\else
  \newcommand{\added}[2][]{#2}
  \newcommand{\deleted}[2][]{}
  \newcommand{\replaced}[3][]{#2}
\fi

\begin{document}

\title{When Seconds Count: Designing Real-Time VR Interventions for Stress Inoculation Training in Novice Physicians}


\author{Shuhao Zhang}
\authornote{Both authors contributed equally to this research.}
\email{zhangshh12024@shanghaitech.edu.cn}
\orcid{0009-0008-1933-1869}
\affiliation{
\institution{School of Information Science and Technology, ShanghaiTech University}
  \city{Shanghai}
  \country{China}
}
\affiliation{
  \institution{Epitome}
  \city{Shanghai}
  \country{China}
}

\author{Jiahe Dong}
\authornotemark[1]
\email{dongjh@shanghaitech.edu.cn}
\orcid{0009-0003-1537-0717}
\affiliation{
\institution{School of Information Science and Technology, ShanghaiTech University}
  \city{Shanghai}
  \country{China}
}

\author{Haoran Wang}
\email{hw3180@nyu.edu}
\orcid{0009-0005-3254-0604}
\affiliation{
  \institution{Interactive Media Arts, New York University Shanghai}
  \city{Shanghai}
  \country{China}
}
\affiliation{
  \institution{Epitome}
  \city{Shanghai}
  \country{China}
}

\author{Chang Jiang}
\email{cjiang_fdu@yeah.net}
\orcid{0000-0002-7468-3372}
\affiliation{
  \institution{Shanghai Clinical Research and Trial Center, ShanghaiTech University}
  \city{Shanghai}
  \country{China}
}

\author{Quan Li}
\authornote{Corresponding Author.}
\email{liquan@shanghaitech.edu.cn}
\orcid{0000-0003-2249-0728}
\affiliation{
\institution{School of Information Science and Technology, ShanghaiTech University}
  \city{Shanghai}
  \country{China}
}
\affiliation{
  \institution{Yibandao (Suzhou) Intelligent Technology Co., Ltd.}
  \city{Suzhou}
  \country{China}
}

\renewcommand{\shortauthors}{Shuhao Zhang and Jiahe Dong, et al.}

\begin{abstract}
Surgical emergencies often trigger acute cognitive overload in novice physicians, impairing their decision-making under pressure. Although Virtual Reality–based Stress Inoculation Training (VR-SIT) shows promise, current systems fall short in delivering real-time, effective support during moments of peak stress. To bridge this gap, we first conducted a formative study (N=12) to uncover the core needs of novice physicians for immediate assistance under acute stress and identified three key intervention strategies: self-regulation aids, procedure guidance, and emotional/sensory support. Building on these insights, we designed and implemented a novel VR-SIT system that incorporates a just-in-time adaptive intervention framework, dynamically tailoring support to learners' cognitive and emotional states. We then validated these strategies in a user study (N=26). Our findings provide empirical evidence and design implications for next-generation VR medical training systems, supporting physicians in sustaining cognitive clarity and accurate decision-making in critical situations.
\end{abstract}




\begin{teaserfigure}
  \centering
  \begin{subfigure}[b]{0.55\textwidth}
      \centering
      \includegraphics[height=3.5cm]{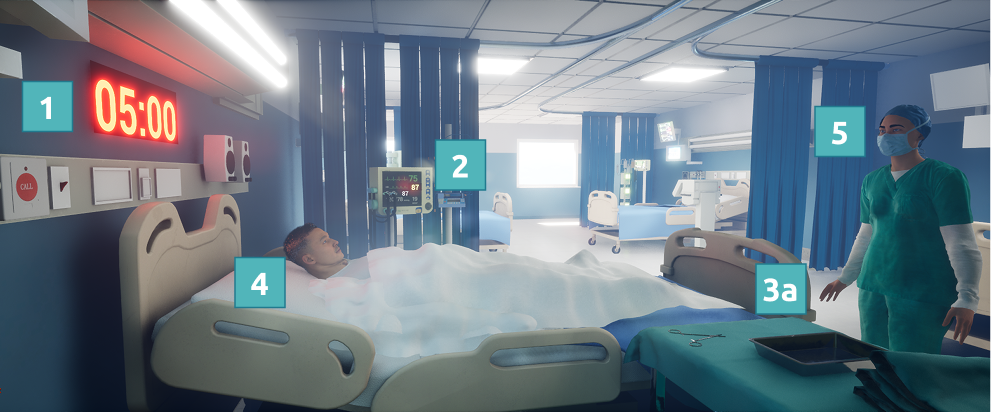}
      \caption{First-person view of the virtual ward.}
      \label{fig:teaser_left}
  \end{subfigure}
  \hfill
  \begin{subfigure}[b]{0.44\textwidth}
      \centering
      \includegraphics[height=3.5cm]{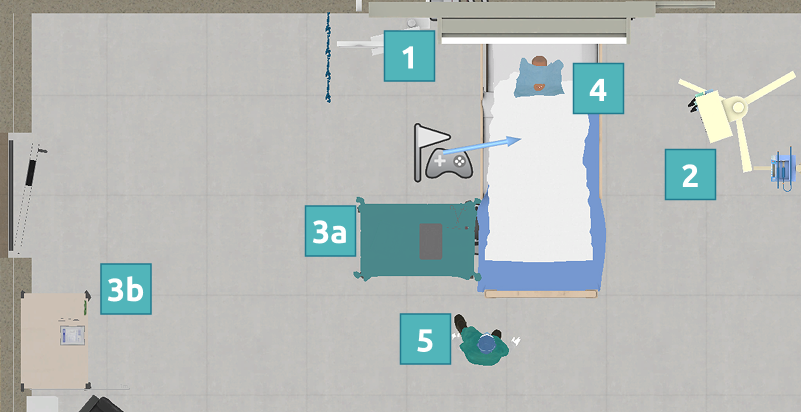}
      \caption{Top-down view of the ward.}
      \label{fig:teaser_right}
  \end{subfigure}
  \caption{In this VR environment, the trainee can undergo Stress Inoculation Training in a high-fidelity simulated setting to learn techniques for coping with acute, sudden stress. Numbers 1–5 indicate important entities in the simulation: 1) the call bell, surgical lamp and time indicator used to start and end task sequences; 2) the patient monitor displaying heart rate and oxygen saturation; 3) the instrument table for placing critical instruments (3a) and the supply table (3b); 4) the virtual patient; 5) the nurse NPC.}
  \Description{A two-panel figure showing a virtual reality hospital ward used for stress inoculation training. Panel (a) displays a first-person view from the trainee's perspective, showing a virtual patient in a bed, a nurse NPC, and surrounding medical equipment. Panel (b) shows a top-down schematic view of the ward, illustrating the layout of the room and the placement of the entities. Numbered labels from 1 to 5 in both panels identify key components as detailed in the caption: 1, the call bell, surgical lamp and timer; 2, the patient monitor; 3, the instrument table and supply table; 4, the virtual patient; and 5, the nurse NPC.}
  \label{fig:teaser}
\end{teaserfigure}

\definecolor{myFrameColor}{HTML}{FF0000}
\definecolor{myBgColor}{HTML}{FFBBBB}
\definecolor{iconfillcolor}{HTML}{51B5BA}
\definecolor{iconbordercolor}{HTML}{217A8E}

\newcommand{\Ticon}[1]{
  \begingroup
  \setlength{\fboxsep}{2pt}
  \setlength{\fboxrule}{1pt}
  \fcolorbox{myFrameColor}{myBgColor}{\textbf{#1}}
  \endgroup
}

\newcommand{\capsulebw}[1]{
  \begingroup
  \setlength{\fboxsep}{3pt}
  \setlength{\fboxrule}{0.5pt}
  \fcolorbox{cyan!50!black}{cyan!10}{
    \sffamily\bfseries\footnotesize #1
  }
  \endgroup
}

\newcommand{\teasericon}[1]{
  \begingroup
  \setlength{\fboxsep}{3pt}
  \setlength{\fboxrule}{1pt}
  \fcolorbox{iconbordercolor}{iconfillcolor}{
    \color{white}\normalsize\bfseries #1
  }
  \endgroup
}
\maketitle
\section{Introduction}

\par Surgical emergencies, as a core component of medical practice, often induce substantial psychological stress on novice physicians due to their inherent complexity and high-stakes nature~\cite{christakis2012measuring, tat2025burnout}. This pressure is especially acute in scenarios involving postoperative complications or device malfunctions, where clinicians face not only intense performance anxiety but also severe time constraints~\cite{tam2025effects}. Unlike routine stressors, these events often arise unexpectedly, demanding immediate responses with limited cognitive preparation — a typical characteristic of acute stress. In such high-pressure, time-sensitive environments, physicians are prone to cognitive disruptions characterized by fear, anxiety, and uncertainty—a phenomenon commonly described in cognitive psychology as ``choking under pressure''~\cite{baumeister1984choking}. Studies have demonstrated that anticipatory stressors, such as fear of litigation or potential harm to professional reputation, can further intensify this cognitive overload, leading to decision paralysis and compromising patient safety~\cite{leblanc2009effects,grantcharov2019acute}. Consequently, developing training methods that mitigate cognitive disruptions is critical for both physician well-being and patient outcomes. Addressing this challenge—enabling novice physicians to retain cognitive clarity and make timely, accurate decisions under extreme conditions—has become a pressing concern in both medical education and the field of HCI~\cite{elendu2024impact,mache2018mental}.

\par Conventional medical training methods—such as classroom lectures, case-based discussions, and simulation exercises—are effective in delivering theoretical knowledge but largely depend on passive learning strategies. These approaches often lack situational realism and timely feedback, making it challenging for trainees to translate conceptual understanding into effective action during high-pressure clinical situations~\cite{langote2024human}. To overcome these limitations, medical education has increasingly adopted techniques inspired by cognitive-behavioral therapy. Notable among these are Stress Exposure Training (SET)~\cite{driskell1998stress}, which emphasizes performance under pressure through direct exposure, and Stress Inoculation Training (SIT)~\cite{meichenbaum1985stress}, a phased method where learners first acquire and rehearse coping mechanisms before applying them in stressful contexts. However, implementing these paradigms in medical training poses considerable challenges. Simulating high-risk clinical emergencies for training purposes faces significant logistical, financial, and ethical constraints~\cite{solli2022alternating}. In this regard, virtual reality (VR) offers a promising alternative by allowing trainees to engage with immersive, high-stakes scenarios in a repeatable, safe, and controlled environment. VR aligns well with the principles of SET and SIT, enabling stress exposure and skill rehearsal in ways that are both scalable and adaptable to real-world demands~\cite{weiss2023don,blanchard2024combining}. While both paradigms contribute valuable perspectives, SIT's structured, multi-phase design that explicitly incorporates the acquisition and rehearsal of coping strategies provides a more comprehensive theoretical foundation for developing training systems with active, real-time support~\cite{nehra2014comparative}. In contrast, SET focuses primarily on habituation through repeated exposure. Consequently, our research adopts the SIT paradigm, examining how VR can enhance its application, particularly through in-simulation guidance and adaptive feedback mechanisms.

\par However, despite VR's theoretical potential, current implementations fall short in addressing key challenges during acute cognitive overload. A review of the existing literature reveals two critical research gaps. The first, the \textbf{Paradigm Gap (G1)}, arises from the focus of most VR-based Stress Inoculation Training (VR-SIT) systems on stress induction and post-hoc evaluation. This approach largely overlooks the need for real-time cognitive and decision-making support when trainees are experiencing cognitive overload and are at risk of mental breakdown. For instance, trainees often lack in-simulation guidance when faced with critical decisions and hesitation. The second, the \textbf{Strategy Gap (G2)}, pertains to stress intervention methods in HCI. While research in this area is emerging, many of these strategies are time-consuming and not well-suited to the rapid-response nature of surgical emergencies. The question of which intervention techniques can deliver real-time, effective support in such high-stake contexts remains unresolved~\cite{stuart2020applying,raggi2025inducing}.

\par To address \textbf{G1} and \textbf{G2}, our study first explored the design space for customizing intervention strategies tailored to time-critical scenarios in surgical emergencies. This exploration led to our first research question: \textit{RQ1: What forms of real-time support do novice physicians require to manage acute cognitive overload effectively during high-stakes surgical emergencies?} To answer \textbf{RQ1}, we conducted a two-tiered focus group study involving both novice and senior physicians. Our findings revealed three core areas for real-time support: \textit{self-regulation aids}, \textit{procedure guidance}, and \textit{emotional/sensory support}. Expert validation of these findings revealed a critical tension between providing immediate assistance and promoting trainee independence. This challenge in designing a balanced, integrated system inspired our second research question: \textit{RQ2: How can the traditional VR-SIT paradigm be adapted to effectively integrate and deliver these diverse real-time interventions, supporting trainees during moments of acute stress?} In response to \textbf{RQ2}, we designed and implemented a novel VR-SIT system based on the Just-In-Time Adaptive Intervention (JITAI) framework~\cite{nahum2016just}. This system operationalizes the three identified support strategies, dynamically delivering them according to the trainee's real-time cognitive and emotional states during a ``post-thyroidectomy neck hematoma'' simulation. The development of this integrated and adaptive platform prompted a critical evaluative question, leading to our final research question: \textit{RQ3: How does the system influence trainees' task performance, cognitive recovery, and subjective experience under acute stress?} To answer \textbf{RQ3}, we conducted a user study with 26 participants to validate the system's effectiveness. The key contributions of this study are threefold:
\begin{itemize}
    \item We conducted a formative study to articulate the design space of time-sensitive stress recovery, identifying three core intervention strategies: \textit{self-regulation aids}, \textit{procedure guidance}, and \textit{emotional/sensory support}.
    \item We designed and implemented a VR-based SIT system that operationalizes these intervention principles, providing a research platform for the empirical evaluation of their effectiveness.
    \item Through a user study involving 26 participants, we provide empirical evidence supporting the effectiveness of the proposed interventions, offering actionable design implications for future VR-SIT systems that aim to deliver timely and effective support during critical moments of stress.
\end{itemize}


\section{Related Work}

\par This study lies at the intersection of medical education, cognitive psychology, and human-computer interaction. In this section, we review the relevant literature to establish the theoretical foundation for our research. We begin by examining key stress management training paradigms—\textit{Stress Exposure Training} and \textit{Stress Inoculation Training}—and their applications. Next, we explore how VR has emerged as a modern platform for these paradigms and analyze the limitations of current VR-based implementations. Finally, we discuss related intervention strategies from the HCI field, identifying critical opportunities and gaps that will guide the design and direction of this study.

\subsection{Stress Management Paradigms}
\par The deleterious effects of acute stress on cognition, often termed ``choking under pressure''~\cite{baumeister1984choking}, are well-documented and particularly critical in high-stakes surgical emergencies~\cite{grantcharov2019acute,tam2025effects}. Cognitive psychology attributes this phenomenon to the depletion of limited working memory resources under stress~\cite{leblanc2009effects}, which in turn impairs the executive functions essential for clinical decision-making. Specifically, stress narrows attentional focus, leading to missed yet crucial peripheral cues, and shifts reasoning from deliberate, analytical processes to fast, error-prone heuristics~\cite{staal2004stress,eysenck2007anxiety,croskerry2003importance}. These findings underscore the need for training approaches that explicitly address performance under pressure. To this end, cognitive-behavioral science offers two foundational paradigms to address this challenge.

\par The first, \textbf{SET}, is based on the principle of habituation~\cite{driskell1998stress,grissom2009habituation}. This paradigm involves repeatedly exposing trainees to stress-inducing environments, allowing their physiological and psychological responses gradually diminish, thereby improving their performance under similar stressful conditions in the future~\cite{driskell2001does,driskell2018stress}. In contrast, \textbf{SIT}, grounded in cognitive-behavioral therapy, actively enhances an individual's coping abilities through a structured teaching process~\cite{beck2020cognitive,meichenbaum1983stress}. SIT typically involves three phases: the \textit{cognitive preparation} phase, the \textit{skill acquisition \& rehearsal} phase, and the \textit{application and practice} phase~\cite{prachyabrued2019toward}. 

\par \replaced{Crucially, the efficacy of SIT rests on developing metacognitive awareness, i.e., the executive capacity to monitor, evaluate, and regulate one's own cognitive processes~\cite{flavell1979metacognition}. In contrast to SET, which focuses on physiological habituation through repeated exposure~\cite{driskell1998stress,long1984aerobic,foa1999comparison}, SIT emphasizes cultivating an internal supervisory mechanism that enables trainees to recognize and manage their stress responses. During the preparation phase, trainees acquire \textit{metacognitive knowledge} by learning how stress affects cognition and performance. The skill acquisition phase then transforms this knowledge into \textit{metacognitive regulation strategies} aimed at addressing specific cognitive vulnerabilities. In the application phase, trainees engage in \textit{metacognitive monitoring} in real-time, identifying early signs of performance decline and intentionally applying countermeasures to maintain executive control. This progression, from tolerating stress to actively regulating its cognitive impact, is central to SIT~\cite{meichenbaum1983stress}, particularly in complex medical emergencies where sustaining decision-making authority is essential.}{This systematic, multi-phase approach, which teaches and reinforces coping skills, distinguishes SIT from SET, which focuses primarily on exposure\mbox{~\cite{long1984aerobic,foa1999comparison}}.}

\par While both paradigms have proven effective, their implementation in traditional medical education—using tools like high-fidelity manikins and standardized patients~\cite{higham2020simulation,bradley2006history,gerzina2019standardized}—faces several critical challenges. These approaches provide safe, instructor-led environments~\cite{krishnan2017pros}, but inducing comparable levels of psychological stress across sessions remains difficult, affecting training reliability~\cite{ignacio2015comparison,rutherford2019systematic}. Furthermore, assessments are often reliant on instructors' subjective evaluations, lacking objective, real-time metrics~\cite{zhen2021learning,carey2020high}, and the high costs of equipment and personnel limit scalability~\cite{rutherford2019systematic,hanshaw2020high}.

\subsection{VR for SIT}

\par Virtual Reality has emerged as a compelling platform for overcoming the challenges associated with stress management training. Its ability to deliver immersive, repeatable, and cost-effective training environments, along with precise experimental control, makes VR an ideal medium for operationalizing stress management paradigms, particularly SIT~\cite{garcia2001redefining,muhling2023virtual,neher2025virtual}. Empirical research has demonstrated VR's capacity to induce anxiety and emotional responses comparable to real-life situations, thus providing a strong foundation for the effective application of stress training~\cite{stinson2014feasibility,rizzo2012strive}.

\par However, while SIT's theoretical framework is comprehensive, existing VR systems predominantly focus on the \textit{application and practice phase} of SIT during their implementation and exhibit notable limitations. Foundational research, such as the work by ~\citet{prachyabrued2019toward}, has successfully demonstrated that VR can create valid stress-inducing scenarios for emergency personnel, concentrating primarily on stress exposure and its psycho-physiological impacts. While this line of research is crucial for validating VR as a training platform, it highlights a significant gap: the primary focus has been on stress induction and post-hoc evaluation, rather than integrating real-time mechanisms that provide skill guidance or cognitive assistance during critical moments~\cite{blanchard2024combining}. As a result, current VR-SIT implementations largely neglect the \textit{skill acquisition \& rehearsal phase} that is fundamental to SIT. This lack of real-time, in-scenario cognitive support forms the basis of \textbf{G1}, which we identify and address in this study.

\par To address this gap, we propose that VR-SIT systems should go beyond exposure and assessment and actively incorporate real-time intervention mechanisms. To explore this design space, we introduce the JITAI framework~\cite{nahum2016just}, which was initially developed for the mobile health field. This framework focuses on identifying moments when users experience cognitive vulnerability under stress and proactively delivering tailored, real-time support. This principle aligns well with the goals of the \textit{application and practice phase} of SIT, enabling trainees to practice and internalize coping strategies at the most stressful points in real-world scenarios~\cite{nahum2015building,henry2025just}. Therefore, this study aims to investigate the manifestation and effectiveness of real-time interventions based on the JITAI framework in high-risk, time-sensitive medical training environments, addressing the current gap in real-time intervention strategies within VR-SIT implementation models.

\subsection{Real-time Intervention in HCI}

\par To operationalize the conceptual model into a functional system, it is essential to review established methods for delivering real-time adaptive interventions. To this end, we turn to the field of HCI, which offers a robust foundation in developing systems capable of sensing user states and providing timely support. In this section, we begin by reviewing sensing technologies for inferring user states, followed by an analysis of the design space of existing intervention strategies.

\subsubsection{\replaced{From User State Awareness to Cognition-Aware Computing}{User State Awareness: The Prerequisite for Intervention}}
\label{sec:2.3.1}
\par A foundational prerequisite for any real-time intervention system is the ability to accurately sense or infer a user's internal state to determine the optimal moment for support~\cite{nahum2016just}. The HCI community has made significant strides in this area, developing a range of techniques that can be broadly categorized into \textit{physio-computing} and \textit{behavioral computing}~\cite{baig2019survey,fairclough2017physiological,scherer2012generic,pantic2006human}. Physio-computing involves the use of sensors to capture physiological signals closely correlated with stress and cognitive load, such as heart rate variability (HRV)~\cite{bhoja2020psychophysiological,sroufe1971effects} and galvanic skin response (GSR)~\cite{turankar2013effects}. In contrast, behavioral computing relies on analyzing user interaction patterns—such as gaze trajectories~\cite{kiefer2017eye}, response latency~\cite{beatty1982task}, and error rates~\cite{just2013intensity}—to infer cognitive states indirectly~\cite{wickens2008multiple}. 

\added{However, detecting raw physiological signals alone is insufficient for high-stakes training; the system must also interpret the user's cognitive context. This has given rise to \textit{Cognition-Aware Computing}, which seeks to infer cognitive processes—such as attention, workload, and alertness—from sensor data to create adaptive systems that respond to the user's mental state~\cite{bulling2014cognition}. Research by \citet{dingler2017building} demonstrated that cognitive alertness fluctuates significantly, highlighting the need for systems capable of detecting these shifts in real-time to provide timely assistance. Additionally, \citet{sarsenbayeva2019measuring} emphasized that internal states, such as stress, can function as ``situational impairments'', similar to environmental noise, which detract from interaction performance. Drawing on these insights, our system leverages physiological data to infer critical moments of cognitive vulnerability, triggering JITAI to restore cognitive capacity.}

\subsubsection{The Design Space of Interventions and Its Critical Contextual Gap}
\label{sec:2.3.2}
\par Once the need for support is identified, \replaced{the design of intervention strategies can be analyzed through the lens of Gross's \textit{Process Model of Emotion Regulation}, which categorizes strategies based on when they impact the emotion-generative process~\cite{gross2015emotion,slovak2023designing}. Reviewing the HCI literature, we identify three prominent approaches mapping to this model.}{the HCI literature offers a broad and nuanced design space of real-time intervention strategies\mbox{~\cite{gross2015emotion,slovak2023designing}}. To systematically examine this space and uncover the underlying design assumptions, we thematically categorize interventions into three prominent approaches.} The first, \textit{cognitive and physiological self-regulation}, involves guiding users through structured exercises or providing granular biofeedback to promote self-awareness and control~\cite{fang2025social,sun2020fostering}. The second, \textit{environmental displacement}, seeks to remove users from the source of stress by immersing them in calming virtual environments, such as natural or abstract landscapes~\cite{ng2023virtual, chandrasiri2020virtual,miller2023awedyssey}. The third, \textit{affective externalization}, focuses on translating emotional states into physical or digital artifacts that users can manipulate or discard to achieve emotional relief~\cite{grieger2021trash}.

\par While diverse in modality, these approaches share a fundamental underlying assumption: \textit{task interruptibility}. Whether it involves pausing to complete a breathing exercise, redirecting attention to an immersive environment, or engaging with an emotional artifact, each strategy presumes the user's ability to momentarily disengage from their main activity. This assumption holds in many common contexts, such as workplace wellness programs or casual interactive systems. However, it breaks down entirely in high-stakes, time-critical domains such as simulated surgical emergencies. Here, the primary task is both cognitively demanding and non-interruptible; any lapse in attention could result in clinical deterioration or virtual patient death. Consequently, conventional pause-dependent interventions are not only impractical but potentially harmful.

\par This disconnect reveals a critical gap: \textbf{G2}. In high-risk, time-sensitive environments where uninterrupted task performance is essential, there is little empirical understanding of what forms of intervention are both feasible and effective. Addressing this gap requires rethinking intervention strategies: they must operate peripherally, be processed implicitly, and integrate seamlessly with the ongoing workflow without disrupting task execution. This study therefore centers on identifying and empirically evaluating such context-appropriate strategies for real-time intervention in high-stakes simulated settings.

\section{Formative Study}
\label{sec:Formativestudy}

\par This section builds upon the two research gaps identified in previous sections: current VR training systems fail to fully leverage the potential of SIT, and commonly used HCI interventions are not well-aligned with the continuity and urgency of real-world surgical emergencies. To begin addressing these challenges, our goal is to design a VR-SIT system that serves as a platform for exploring the design space of real-time stress interventions. As a critical first step, this formative study aims to identify the core stressors that should be simulated in training, as well as the types of support novice physicians need when facing such stressors, thereby addressing \textbf{RQ1}.

\par
\begin{figure*}[t]
    \centering
    \includegraphics[width=\textwidth]{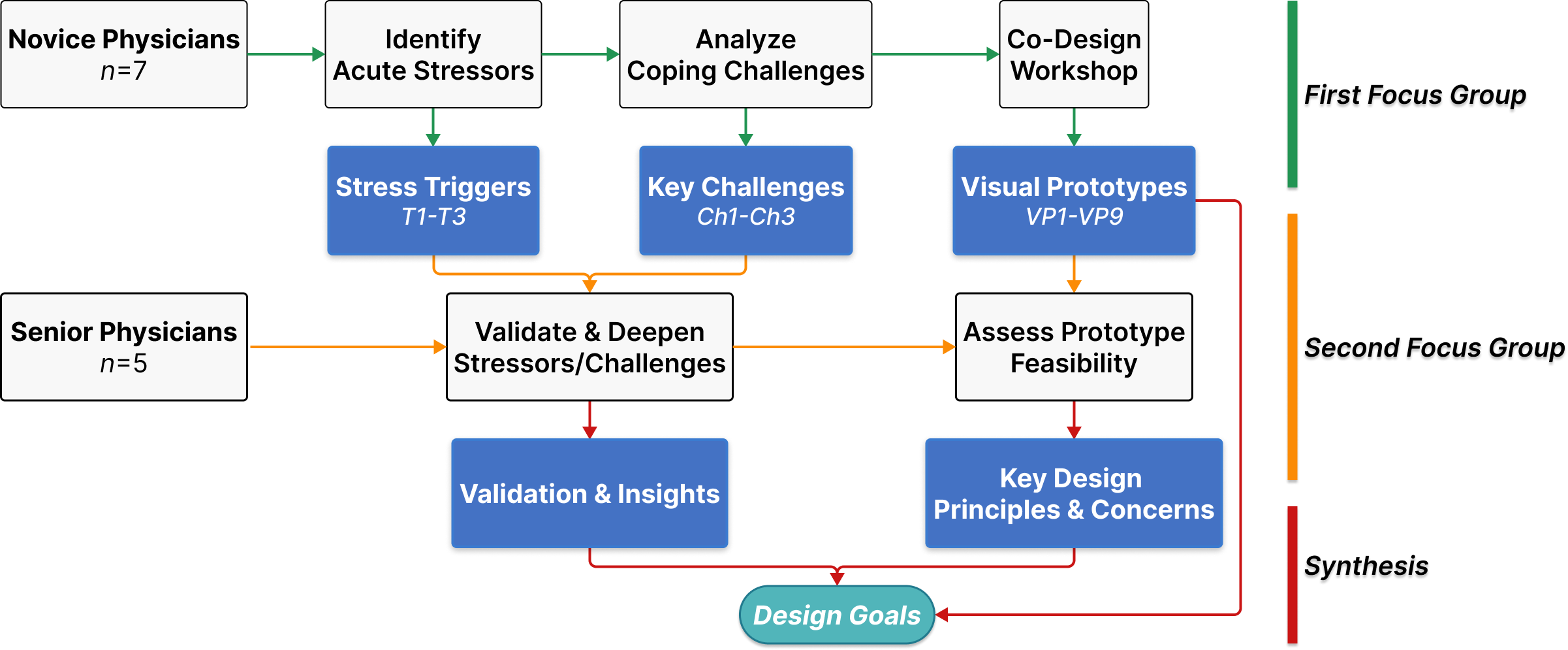}
    \caption{Procedure of formative study.}
    \Description{A flowchart diagram titled "Fig. 2. Procedure of formative study." illustrates a three-phase research process: First Focus Group, Second Focus Group, and Synthesis. The process begins with "Novice Physicians (n=7)" who participate in the "First Focus Group." This involves three steps: "Identify Acute Stressors," leading to "Stress Triggers (T1-T3)"; "Analyze Coping Challenges," resulting in "Key Challenges (Ch1-Ch3)"; and a "Co-Design Workshop," which produces "Visual Prototypes (VP1-VP9)." The second phase involves "Senior Physicians (n=5)" in the "Second Focus Group." They first "Validate \& Deepen Stressors/Challenges," using the inputs from the novice physicians' findings, which generates "Validation \& Insights." They then "Assess Prototype Feasibility," based on the visual prototypes, leading to "Key Design Principles \& Concerns." The final "Synthesis" phase shows that all outputs—Stress Triggers, Key Challenges, Visual Prototypes, Validation \& Insights, and Key Design Principles \& Concerns—feed into the final outcome, an oval labeled "Design Goals." The flow is indicated by colored arrows: green for the first focus group, orange for the second, and red for the synthesis phase.}
    \label{fig:formative_study}
\end{figure*}

 To achieve this, we employed a two-tiered focus group methodology~\cite{krueger2014focus}. This approach enabled us to first develop a bottom-up understanding of novice physicians' lived experiences and needs, and subsequently apply top-down validation from clinical experts to refine these needs into safe and effective intervention concepts. This section outlines the study's methodology and key findings, beginning with a description of the study setup and participant recruitment. We then present insights from the initial focus group with novice physicians, followed by the co-design process and refinement of strategies in the second focus group with senior clinicians. Finally, we synthesize these insights into a set of actionable design goals that will inform the development of our VR system. ~\autoref{fig:formative_study} illustrates the process of formative study.

\begin{table}[h]
    \centering
    \caption{Participant Demographics.}
    \label{tab:participant-info}
    \begin{tabular}{@{}llllll@{}}
        \toprule
        ID & Training Level & Exp. & Department & Gen. & Group \\ \midrule
        N1 & Resident & 2 yr & Emergency Med. & M & Novice \\
        N2 & Resident & 2 yr & ICU & F & Novice \\
        N3 & Resident & 1 yr & Internal Med. & M & Novice \\
        N4 & Resident & 1 yr & Emergency Med. & F & Novice \\
        N5 & Resident & 1 yr & Gen. Surgery & M & Novice \\
        N6 & Med Stu. & <1 yr & Clinical Rot. & F & Novice \\
        N7 & Med Stu. & <1 yr & Clinical Rot. & F & Novice \\
        S1 & - & 15 yr & Orthopedic Surg. & M & Senior \\
        S2 & - & 12 yr & Gen. Surgery & F & Senior \\
        S3 & - & 12 yr & ENT & F & Senior \\
        S4 & - & 11 yr & ENT & M & Senior \\
        S5 & - & 10 yr & Orthopedic Surg. & M & Senior \\ \bottomrule
    \end{tabular}
    \vspace{-10pt}
\end{table}

\subsection{Setup}

\par \replaced{We conducted a two-tiered focus group study involving $12$ participants. Recruitment was conducted through posters distributed in partner hospitals and via snowball sampling. The study was approved by our Institutional Review Board, and all participants provided written informed consent. A detailed summary of the participant demographics can be found in\mbox{~\autoref{tab:participant-info}}.}{We conducted a two-tiered focus group study with a total of $12$ participants recruited from local hospitals, as illustrated in\mbox{~\autoref{fig:formative_study}}. The study received approval from our Institutional Review Board, and all participants provided written informed consent. Detailed demographics are summarized in\mbox{~\autoref{tab:participant-info}}.}

\replaced{The first tier ($N_{novice}=7$) focused on mapping the problem space. We operationally defined ``novice physicians'' as residents or medical students with less than three years of clinical experience, consistent with the residency training standards of the participating hospitals. This session explored their lived experiences, cognitive challenges, and unmet needs working in high-stress environments. Insights from this group, including user-generated visual prototypes, informed the design direction for the second tier. In the second tier ($N_{senior}=5$), we engaged senior physicians (attending surgeons with over 10 years of clinical practice) to evaluate and refine the initial prototypes, ultimately shaping the system's design goals. Each focus group was co-moderated by two researchers and lasted approximately 75 minutes. Data were collected through audio recordings, transcriptions, and whiteboard annotations. All participants received a \$20 compensation for their time.}{The first tier ($N_{novice}=7$) focused on mapping the problem space by exploring the lived experiences, cognitive challenges, and unmet needs of novice physicians working in high-stress environments. Insights from this group, including user-generated visual prototypes, informed the design direction for the second tier. In the second tier ($N_{senior}=5$), we engaged senior physicians with substantial clinical experience to evaluate and refine the initial prototypes, ultimately shaping the system's design goals. Each focus group was co-moderated by two researchers and lasted approximately 75 minutes. Data were collected through audio recordings, transcriptions, and whiteboard annotations. All participants received a \$20 compensation for their time.}

\subsection{First Focus Group: Novice Physicians}
\par To explore novice physicians' stress experiences, coping failures, and concrete needs for real-time support in high-risk, time-critical surgical emergencies, we conducted the first round of focus groups. This phase was designed to surface key pain points to be prioritized in the system design and to establish a foundation for subsequent prototype development. The session involved seven participants: five resident physicians (N1–N5) and two graduating medical students (N6-N7), all of whom had prior clinical rotation experience. The discussion was organized into three stages: identifying acute stressors, analyzing coping breakdowns, and engaging in a co-design workshop focused on potential intervention strategies.

\subsubsection{Stage 1: Identification of Acute Stressors}
\par In the first stage, participants were invited to recall and share real-life experiences of unexpected surgical emergencies encountered during clinical rotations or internships, with the goal of identifying specific triggers of ``acute stress''. We explicitly distinguished this form of stress from ``chronic stress'' typically observed in routine nursing contexts, emphasizing characteristics such as sudden onset, disruption of procedural workflows, high clinical risk, and the need for immediate response.

\par Participants described a range of high-pressure scenarios, including intraoperative patient hypoxia, uncontrolled bleeding, equipment malfunctions, inability to locate critical instruments, and lack of timely support. These situations collectively define a high-stakes, time-sensitive clinical domain marked by concentrated responsibility and elevated cognitive demand. As N1 noted: ``\textit{What is taught in books and classrooms are standard procedures, but the real situation is much more complex.}'' N6 added: ``\textit{Although we were taught how to handle potential complications, their likelihood seemed so low. When you're standing in front of a patient, you might not remember all the treatment protocols.}''

\par We identified three primary categories of acute stress triggers reported in these scenarios:

\par \Ticon{T1}\textbf{Environment-related Triggers.} These stressors arise from the external environment rather than direct clinical procedures. Although these factors do not directly affect clinical decision-making, they can significantly disrupt procedural efficiency and safety. Typical situations include equipment failure, shortage of supplies, or inadequate collaborative resources. As N4 described: ``\textit{You're ready to suture a blood vessel, but you can't find a suitable thread even after searching the entire table. In that moment, every second counts, yet you still have to call a nurse to bring another set.}'' Such failures at critical moments interrupt procedural flow, provoke anxiety about external support, and intensify time pressure, especially for novice physicians who are highly likely to experience significant psychological burden.

\par \Ticon{T2}\textbf{Process-related Triggers.} These stressors occur during the clinical procedures, typically characterized by sudden patient condition changes, requiring physicians to complete assessment, judgment, and intervention decisions within an extremely short time frame, representing a classic high-task intensity scenario. N2 recounted a surgical episode: ``\textit{Everything was fine at the beginning, but suddenly the patient's blood pressure just dropped. I was totally stunned and kept thinking: What should I do?}'' Such sudden changes disrupt operational expectations, highlighting decision-making pressure and risk responsibility, with many novice physicians experiencing significant stress and psychological impact in such situations.

\par\Ticon{T3}\textbf{Outcome-related Triggers.} These stressors do not stem directly from external environments or sudden condition changes, but emerge when the operation's results do not meet expectations, often triggering self-doubt and frustration in the performer. As N7 recalled: ``\textit{I once tried to draw blood from a patient and failed three times. I did exactly what the instructor had taught me, but there was no blood. The patient grew increasingly anxious, and my hands started shaking.}'' Even without apparent errors, when operational results deviate from expectations, novice physicians tend to attribute this to personal incompetence, thus forming a deeper psychological burden.

\subsubsection{Stage 2: Analysis of Stress Responses and Coping Challenges}

\begin{figure*}[t]
    \centering
    \includegraphics[width=\textwidth]{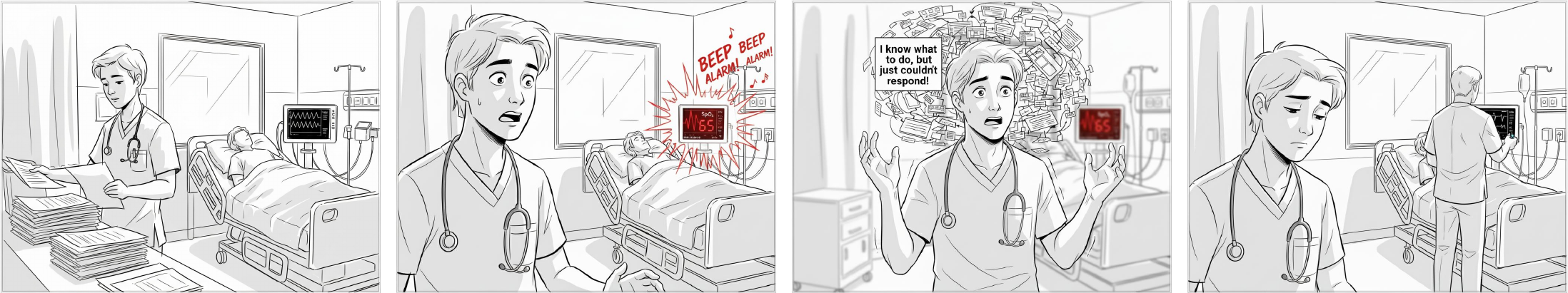}
    \caption{One storyboard illustrating \textbf{Ch1}: Cognitive Freeze.}
    \Description{A four-panel, black-and-white storyboard illustrating the concept of "Cognitive Freeze", which is our challenge 1. Panel 1: A male nurse is calmly reviewing patient charts in a hospital room. In the background, a patient is resting in a bed connected to a vital signs monitor showing stable readings. Panel 2: Suddenly, the vital signs monitor is blaring an alarm, and the patient appears to be in distress. The physician looks up from the charts with a shocked and alarmed expression. Panel 3: The physician is shown frozen in place. His head is depicted as a chaotic scribble of tangled lines, symbolizing his mental state. A thought bubble reads, "I know what to do, but just couldn't respond!". Panel 4: The physician stands in the foreground, looking down with a dejected and disappointed expression. In the background, other medical staff have rushed in and are now attending to the patient.}
    \label{fig:Ch}
\end{figure*}

\par After identifying typical high-pressure events, we further explored the subjective experiences of participants. From their feedback, we distilled three key challenges:

\par \textbf{Ch1. Cognitive Freeze.} Many novice physicians experience pronounced cognitive freezing when confronted with sudden, high-stress events. In these moments, they recognize the need to act but are unable to make timely decisions or execute appropriate procedures. This response is particularly evident in situations that deviate from routine protocols, such as equipment malfunctions or abrupt changes in a patient's physiological status. As N3 described: ``\textit{At that moment, it felt like my brain was frozen—I knew what to do, but just couldn't respond.}'' Rather than being a matter of individual difference, this phenomenon reflects a common acute stress response. Participants consistently reported accompanying physiological symptoms—such as increased heart rate, cold extremities, and rapid breathing—which further intensified their psychological strain. These reactions often led to a steep and immediate decline in cognitive functioning. As N6 explained: ``\textit{My mind suddenly became chaotic, as if all my knowledge had vanished, leaving only a sense of instinctive panic.}'' 

\par \textbf{Ch2. Decision Lag.} In addition to cognitive freeze, another prevalent challenge is decision lag, characterized by excessive hesitation and repeated self-verification during procedures, particularly when swift, decisive action is required. Novice physicians often default to over-cautious behavior driven by a strong fear of making mistakes, which significantly impairs execution efficiency. As N2 observed: ``\textit{You become so afraid of making mistakes that you want to confirm every action multiple times, ultimately wasting the most precious time.}'' Similarly, N5 reflected: ``\textit{I would pause and question whether I had done something wrong, even though I had practiced that step many times.}'' This hesitation is not solely a result of insufficient skill. More commonly, it stems from a lack of confidence, the absence of immediate feedback, and the tendency to overthink potential negative consequences. Several participants also mentioned that without clear prompts or a sense of progress, they struggled to assess whether they were ``\textit{"on the right track}''.

\par \textbf{Ch3. Social Isolation.} A recurring theme among participants was the feeling of being ``Isolated and unsupported'', particularly during independent on-call shifts or emergency tasks at night. In such situations, the absence of immediately accessible senior support imposes significant psychological pressure and a profound sense of helplessness on novice physicians. As N1 expressed: ``\textit{The most terrifying part isn't the procedure itself, but being the frontline doctor with no senior staff nearby to turn to—that feeling of loneliness and fear. All the responsibility falls on you.}'' This perceived isolation not only amplifies stress but can also lead to a loss of initiative or even passive inaction. As N7 explained: ``\textit{When I notice something abnormal, my instinct is to look for a teacher. But the problem is, backup takes time to arrive, and until then, you're on your own.}''

\par These challenges not only uncover the core barriers faced by novice physicians in high-pressure scenarios but also provide critical insights that inform the subsequent prototype design. We created storyboards for each challenge to present at the second focus group; \autoref{fig:Ch} shows one example.

\subsubsection{Stage 3: Co-Design Workshop}

\par In the final stage of the focus group, we conducted a prototype co-design workshop, employing low-fidelity design tools such as visual sketches and keyword cards to help participants articulate and externalize their visions for an ideal VR-based support system. Although participants expressed diverse perspectives and priorities, their ideas converged around three core functional directions for potential intervention strategies:
\begin{itemize}
\item \textbf{Enhancing Self-regulation Capabilities}: This category of intervention focuses on helping users regain cognitive control during moments of freezing, thereby restoring a sense of agency. Participants expressed a strong preference for rhythmic, guided UI feedback to anchor their attention and reduce mental chaos. Addressing Ch1, N6 commented, ``\textit{I want a rhythmic visual prompt to pull me out of the chaos.}'' Proposed design ideas include real-time heart rate visualization, stress-level feedback, and guided breathing cues.
\item \textbf{Providing Operational Flow Guidance}: This intervention aims to reduce cognitive load and reinforce procedural rhythm, helping novice physicians stay on track amid chaotic situations. In response to Ch2, participants emphasized the need for timely confirmations and forward-looking task cues. As N2 suggested, ``\textit{It would be best to have a task list like in a game so that I know how many steps are left.}'' Design proposals include step-by-step task checklists and directional guidance elements (e.g., navigation arrows).
\item \textbf{Providing Emotional and Sensory Support}: Aimed at mitigating feelings of isolation and anxiety, this intervention seeks to stabilize users emotionally by modulating environmental stimuli and simulating supportive presence. This directly responds to Ch3. As N4 stated: ``\textit{Sometimes the beeping makes me even more anxious,}'' while N6 added: ``\textit{Even virtual encouragement can make me feel like I'm not fighting alone.}'' Proposed concepts include noise filtering and virtual non-player characters (NPCs) to simulate companionship and reassurance.
\end{itemize}

Across the three functional directions, a total of nine visual prototypes (VP1–VP9) were derived, with three prototypes developed for each direction. Comprehensive descriptions of these prototypes are presented in ~\autoref{appendix:B}, and all associated images are included in the supplementary materials. ~\autoref{fig:VP1} shows two visual prototype diagrams we generated based on these descriptions.


\subsection{Second Focus Group: Senior Physicians}

\par Building on the insights gathered from novice physicians, we conducted a second focus group with the following two objectives: first, to validate and expand upon the previously identified stressors (T1--T3) and challenges (Ch1--Ch3) through the lens of experienced surgical practitioners; second, to evaluate the feasibility of the co-designed visual prototypes and iteratively refine them, thereby shaping the design solution space. This session involved five senior surgical physicians (S1--S5), all of whom had extensive clinical and teaching experience. The discussion was structured in two stages and grounded in materials from the first focus group, including the identified stressors, challenges, and low-fidelity prototype sketches.

\subsubsection{Stage One: Validation and Deepening of Stressors and Challenges}

\par In this stage, we presented the three major stressors and corresponding challenges refined from the first focus group to the participants, and invited the experts to evaluate, validate, and enrich them based on their teaching experience and clinical observations.

\par The experienced physicians generally acknowledged that the identified stressors and challenges closely aligned with real-world clinical experiences, accurately capturing the typical stress responses of novice physicians in emergency surgical situations. As S3 noted: ``\textit{We've all been through these moments. They're not rare exceptions, but scenarios you'll encounter repeatedly throughout your career.}'' With the representativeness of these scenarios confirmed, participants further contributed by deepening the discussion through a mechanistic lens, offering insights into the underlying factors driving these stress responses.

\par \textbf{Elaborating on the Mechanism of Stressors.} The experts emphasized that stressors in clinical settings rarely occur in isolation; instead, they often interact dynamically and exert compounding effects, creating a highly complex and volatile stress environment. For example, an equipment malfunction (T1) may trigger physiological instability in the patient (T2) or result in unanticipated procedural outcomes (T3). As S2 notes: ``\textit{Stressors can snowball. Novices often focus only on the immediate issue, such as bleeding, but may overlook that it stems from a prior operational error.}'' S4 further adds: ``\textit{These stressors tend to accumulate and often interact with underlying chronic stressors, such as lighting, instrument noises, ambient sounds from adjacent areas, or even clinical odors. Together, these factors contribute to a heightened sense of tension, making novice physicians more prone to panic in unexpected situations.}''

\par \textbf{Confirmation and Supplementation of Identified Challenges.} Experts strongly affirmed the presence of ``cognitive freeze'' (Ch1) and ``decision lag'' (Ch2), attributing both primarily to novice physicians' limited psychological preparedness and lack of prior coping experience. As S4 explained: ``\textit{Experienced physicians can respond effectively because we've encountered similar situations many times. Novices, on the other hand, simply don't have enough contingency strategies to draw from.}''

\par Regarding ``social isolation'' (Ch3), experts' opinions diverged. S2, S4, and S5 emphasized the importance of providing appropriate support mechanisms to alleviate novice physicians' anxiety and promote help-seeking behaviors. As S5 noted: ``\textit{The ward isn't composed solely of medical staff, it may also include patients' families, interns, and others. In such an environment, novice physicians often feel uncomfortable speaking up and hesitate to seek support. Experienced physicians can often read the situation and step in when they sense that a novice is struggling to cope.}'' Conversely, S1 and S3 raised concerns about the potential downsides of offering too much external support, arguing that it might hinder the development of independent thinking and self-reliance. As S1 remarked: \textit{Too much external support could actually diminish their capacity for independent thinking.}'' S3 added: \textit{If novices become overly reliant on others, it may foster a dependency mindset, which is ultimately detrimental to their long-term growth.}''

\begin{figure*}[t]
    \centering
    \includegraphics[width=\textwidth]{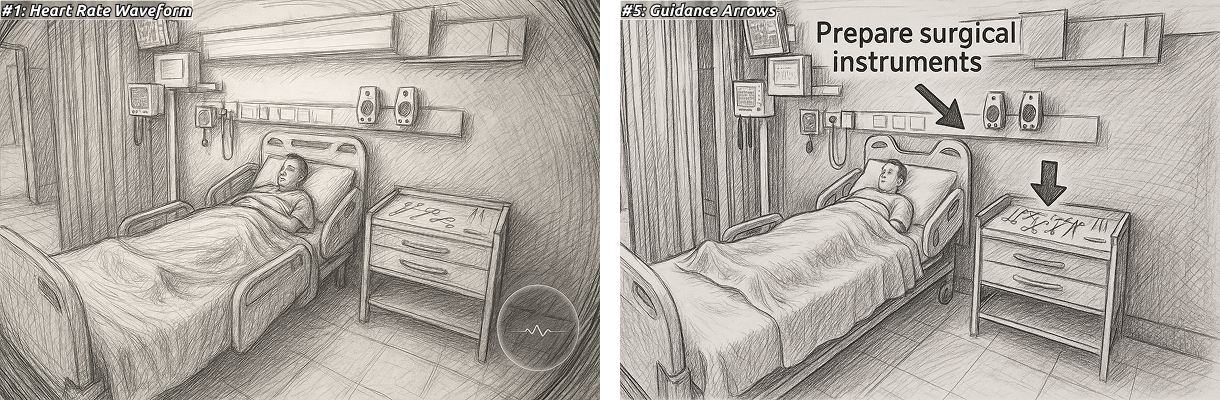}
    \caption{Some outputs from our co-design workshop(VP1\&VP5).}
    \Description{Two side-by-side sketches from a co-design workshop, showing a patient in a hospital bed to illustrate different interface concepts. The left image, titled "#1: Heart Rate Waveform," shows the patient with a graphical overlay of a heart rate waveform near the lower right of the field of view. The right image, titled "#3: Guidance Arrows," shows the same scene but includes the instructional text "Prepare surgical instruments" along with two large arrows pointing to medical equipment on the wall and instruments on the bedside table, demonstrating a visual guidance system.}
    \label{fig:VP1}
\end{figure*}

\subsubsection{Stage Two: Feasibility Assessment of Visual Intervention Prototype}

\par In the second stage, we presented the three major intervention strategies underlying the visual prototypes developed in the first focus group---\textbf{enhancing self-regulation}, \textbf{providing process guidance}, and \textbf{creating emotional support}---to the senior physicians, and facilitated in-depth discussions on their feasibility, potential risks, and opportunities for refinement.

\par \textbf{Self-Regulation Aids.} Self-regulation strategies proposed by novice physicians—such as heart rate visualization and guided breathing prompts—received cautious support from the expert group. While experts agreed that such interventions could help novices disengage from panic and regain cognitive control, they stressed the importance of maintaining minimal interference with core clinical tasks. As S4 remarked when reviewing VP3: ``\textit{An object that suddenly appears at the center of my visual field and triggers automatically? That would startle me and block what I need to see.}'' Following this exchange, the experts recommended implementing breathing guidance as a micro-intervention—a lightweight feature that should either be user-initiated or briefly activated in response to physiological irregularities. For example, a soft, rhythmic pulse might appear subtly at the periphery of the visual field to guide breathing when the system detects elevated heart rate or changes in skin conductance, providing gentle support without interfering with task performance.

\par \textbf{Procedure Guidance.} Experts acknowledged the practicality of prototypes such as task lists and equipment highlighting, which aim to reduce cognitive load and reinforce procedural flow. However, they also expressed concerns that such interventions could inadvertently hinder the development of decision-making. As S3 emphasized: ``\textit{The ultimate goal of training is to cultivate the ability to make independent decisions under acute pressure.}'' S5 added: ``\textit{It's like riding a bicycle with training wheels---but they become overwhelmed once wheels are removed.}'' To address these concerns, the experts proposed a progressive tiered design. Under this approach, the system would remain unobtrusive by default and intervene only when signs of ``decision lag'' (Ch2) are detected. At that point, it would first provide subtle, question-based prompts; if the user confirms the need for support, targeted keyword cues would follow; and only in critical situations would the system escalate to showing detailed operational steps. This tiered strategy aims to balance timely support with the preservation of cognitive autonomy.

\par \textbf{Emotional/Sensory Support.} Expert feedback on this category of prototypes revealed constructive divergence. There was broad consensus on the value of noise filtering (VP8), which was seen as a practical way to reduce environmental cognitive load. However, opinions diverged on the use of a virtual companion (VP9). While some acknowledged its potential to offer reassurance, concerns were raised about the risk of fostering overdependence. As S1 cautioned: ``\textit{Excessive reliance on external approval can undermine the development of independent coping skills—novices ultimately need to learn to take responsibility.}'' S3 echoed this sentiment: ``\textit{I can teach them how to respond, but the goal is for them to make stable decisions even when guidance isn't available.}'' In light of these concerns, the experts recommended shifting from overt emotional support to more implicit, ambient forms of environmental feedback that promote psychological reassurance without diminishing autonomy. 

\par Toward the end of the discussion, experts unanimously emphasized the importance of personalization in system interventions. They advocated for tailoring support strategies to both user characteristics and task attributes. S2 proposed three key variables: ``\textit{First, the learner's personality—whether they are bold or cautious; second, their level of knowledge—whether they know but can't act, or simply don't know; and third, the inherent risk of the procedure.}'' S5 further emphasized: ``\textit{An ideal system should adapt its intervention strategy dynamically, aligning with the user's profile and the situational context.}''

\subsection{Design Goals}

\par To transform the qualitative data from two focus group rounds into actionable design principles, we employed Thematic Analysis~\cite{braun2006using} to systematically synthesize interview recordings, whiteboard notes, and visual sketches. Central to this process was the integration of perspectives from both participant groups: challenges and needs raised by novice physicians (T1-T3, Ch1-Ch3, VP1-VP9) were cross-validated and iteratively refined with feedback from senior physicians. This comparison revealed both consensus and critical divergences between novice requirements and expert insights. From these findings, we distilled five core design goals that address novice physicians' real pain points while embedding SIT and JITAI concepts to ensure safety and effectiveness in clinical practice.

\par \textbf{DG1. Leverage JITAI to deliver \replaced{targeted}{precise}, real-time support during stress episodes.} The system should continuously monitor users' physiological and behavioral indicators, detecting critical moments where stressors  (T1-T3) lead to cognitive freeze (Ch1), decision lag (Ch2), or feelings of social isolation (Ch3). At these intervention-relevant junctures, the system must be capable of promptly initiating targeted support measures, ensuring interventions are both \replaced{targeted}{precise} and timely.

\textbf{DG2. Implement a progressive, tiered support mechanism that balances assistance with learner autonomy.} To address senior physicians' concerns about preserving learners' independence, interventions should follow a tiered progression rather than a uniform approach. By default, the system should refrain from intervening unless prolonged decision latency is detected. It should then escalate support in stages—from subtle prompts or reflective cues to keyword suggestions and, if necessary, explicit operational guidance. This graduated approach ensures critical safety while avoiding over-reliance, fostering learners' capacity for independent clinical reasoning.

\textbf{DG3. Personalize intervention strategies based on individual learner profiles.} 
Given the variability in trainee personalities, experience levels, and cognitive styles, the system should incorporate a personalized intervention framework. By dynamically tailoring support strategies to individual characteristics, it can optimize the relevance and effectiveness of assistance, ensuring that each user receives context-appropriate, learner-specific guidance.

\textbf{DG4. Embed intervention cues seamlessly into the environment to minimize cognitive disruption.} To avoid adding extraneous cognitive load, the system should prioritize subtle, non-intrusive modes of delivering support. Interventions should be ambient and contextually embedded—such as using soft peripheral light pulses for breath pacing or background audio modulation to reduce environmental noise—allowing learners to receive directional guidance without interrupting task flow or clinical immersion.

\textbf{DG5. \replaced{Promote metacognitive awareness conducive to skill internalization.}{Facilitate the internalization and transfer of stress management skills in alignment with SIT principles.}} Beyond supporting task completion, the system should be designed to \replaced{cultivate self-regulatory mechanisms aligned with SIT principles}{cultivate enduring stress coping capabilities aligned with SIT}. Through guided practice, reflection, and feedback, users should progressively internalize techniques for recognizing and regulating stress responses. This will \replaced{build a foundation for potential skill transfer to real-world emergency contexts, rather than merely solving the immediate simulation task}{empower them to manage acute stress in real-world emergency contexts independently, ensuring skill transfer beyond the training environment}.

\section{System Design and Implementation}

\par This section presents the VR-SIT system developed for high-risk, time-critical surgical emergencies. The system addresses \textbf{RQ2} and is guided by the five design goals (\textbf{DG1}–\textbf{DG5}). We first describe the training scenario that provides the contextual background, then outline the supporting technical architecture, and finally detail the three intervention strategies, implemented as modular components with distinct, flexibly applicable functions.

\subsection{Scenario Design}

\subsubsection{Clinical Context and Task Workflow}

\begin{figure*}[t]
    \centering
    \includegraphics[width=\textwidth]{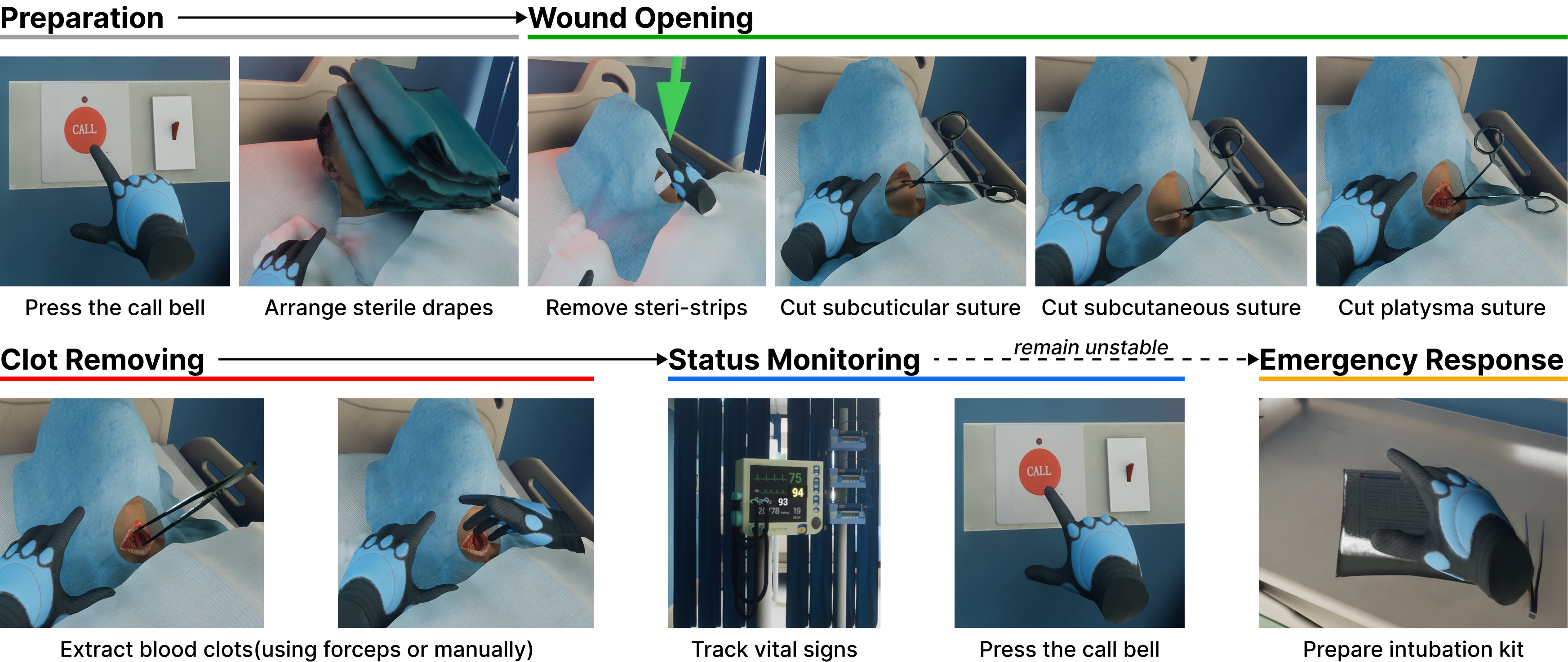}
    \caption{The task workflow.}
    \Description{A workflow diagram illustrating a simulated medical procedure, divided into five stages. The first stage, "Preparation," shows pressing a call bell and arranging sterile drapes. The second stage, "Wound Opening," depicts removing steri-strips, then cutting three types of sutures: subcuticular, subcutaneous, and platysma. The third stage, "Clot Removing," shows the extraction of blood clots with forceps. The fourth stage, "Status Monitoring," displays a vital signs monitor, followed by pressing a call bell. If the patient's status remains unstable, then proceed to the final stage, "Emergency Response," which shows the preparation of an intubation kit. Each step is accompanied by an image from a first-person perspective in a virtual reality environment, showing gloved hands performing the action.}
    \label{fig:Taskflow}
\end{figure*}

\par To ensure both effectiveness and representativeness, it was essential to select a training case that not only reflects real-world clinical challenges but is also suitable for evaluating stress-intervention strategies. Following the second-round focus group, a consensus among senior physicians identified post-thyroidectomy neck hematoma as the core training scenario.

\par Thyroidectomy is a common surgical procedure, with over 150,000 cases performed annually in the United States~\cite{al2016association}. After surgery, patients are typically monitored in general wards where novice physicians, often with limited clinical experience, serve as the primary responders to complications. Post-thyroidectomy neck hematoma, although infrequent (1-2\% incidence), poses a high risk; approximately one-quarter of cases require urgent bedside intervention due to airway compression~\cite{doran2021post, edafe2020reoperation}. Unlike operating room emergencies, these events often occur in resource-limited ward environments without immediate senior support~\cite{iliff2022management}, requiring novice physicians to act independently under extreme time pressure. These features, including sudden onset, severity, time sensitivity, and reliance on autonomous decision-making, make this scenario well-suited for simulation-based research.

\par Experts further emphasized that this scenario mirrors the typical challenges faced by novice physicians in real clinical practice: resource constraints, lack of senior guidance, and error-prone patterns identified in formative studies (e.g., ``\textit{being stunned}'' (S3) or ``\textit{hesitant}'' (S4)). Given that hematoma-induced airway obstruction can become fatal within minutes~\cite{iliff2022management}, novice physicians are often forced to initiate decompression before senior assistance arrives. These conditions capture the essence of high-stakes clinical environments while providing an ideal platform for evaluating intervention strategies.

\par To address such emergencies, standardized protocols like the widely adopted SCOOP guideline~\cite{iliff2022management} are followed. The procedure involves removing the wound dressing, sequentially opening the sutured layers until the incision is exposed, and evacuating the hematoma that compresses the trachea. Since hematomas quickly coagulate into blood clots, clearance often requires forceps or direct manual removal rather than suction alone. Based on these principles, we designed the following core tasks for our VR training scenario, and ~\autoref{fig:Taskflow} shows the implementation in our system.

\begin{enumerate}
\item \textbf{Preparation:} Press the bedside call bell to notify the nursing station and request the crash team, then collect and arrange sterile drapes.
\item \textbf{Wound Opening:} Remove steri-strips and use sterile scissors to cut through skin and subcutaneous sutures layer by layer until the wound is fully exposed.
\item \textbf{Clot Removing:} Extract blood clots using sterile forceps or manual removal.
\item \textbf{Status Monitoring:} Track vital signs via the bedside monitor. If stabilized after clot removal, report via call bell and exit the ward.
\item \textbf{Emergency Response (conditional):} If vital signs remain unstable, request anesthesia team support, prepare the intubation kit, brief the team upon arrival, and hand over before exiting.
\end{enumerate}

\subsubsection{Stress Event Design}\label{sec:stressTriggers}

\par To create a psychologically stressful environment that approximates real emergencies, we implemented a hybrid stress framework combining routine stressors with acute stress events. This design captures the dual influence of persistent baseline pressure and sudden critical challenges. Routine stressors, informed by prior work~\cite{weiss2023don,blanchard2024combining}, include auditory distractions and time constraints: background noise features monitor alarms, intermittent phone rings, and distant conversations, while a visible countdown timer at the patient's bedside indicates the time remaining before the emergency team arrives. These stressors persist throughout training, maintaining continuous tension and contextualizing subsequent interventions.
\par Acute stress events replicate three categories of critical triggers (T1–T3) identified in our formative study, designed to elicit \textit{cognitive freeze} (Ch1) or \textit{decision lag} (Ch2) at pivotal moments, thereby enabling systematic evaluation of intervention strategies. Additionally, the scenario defaults to a solo-duty setting without senior physicians, reinforcing the perception of \textit{social isolation} (Ch3). Concrete implementations are detailed in~\autoref{tab:stressor-info}.

\Ticon{T1} These triggers intentionally disrupt the physical or operational context (e.g., temporary instrument unavailability, fluctuating lighting) to challenge trainees to maintain core task flow amid minor but consequential interruptions. The primary goal is to induce \textit{decision lag} (Ch2) by requiring brief re-orientation and reassessment; in some cases, sudden visibility limitations may also trigger \textit{cognitive freeze} (Ch1).

\Ticon{T2} These triggers introduce abrupt clinical deterioration while compressing available time (e.g., escalated alarms, worsening vitals, shortened countdown). They stress rapid reprioritization and emotional regulation at critical decision points, aiming to provoke both \textit{cognitive freeze} (Ch1) and \textit{decision lag} (Ch2).

\Ticon{T3} These triggers create a mismatch between trainee actions and patient response (e.g., completion of key procedural steps with minimal physiological improvement), prompting hypothesis revision and potential escalation under uncertainty. The intended effect is \textit{decision lag} (Ch2) in a solo-duty setting, which also reinforces perceived \textit{social isolation} (Ch3).

\begin{table*}[t]
    \centering
    \caption{Acute stress triggers: detailed events, timing, and targeted challenges.}
    \label{tab:stressor-info}
    \begin{tabular}{p{1.2cm} p{2.5cm} p{3cm} p{6cm} p{3cm}}
\toprule
\textbf{Category} & \textbf{Event} & \added{\textbf{Timing}} & \textbf{Description} & \textbf{Targeted Challenge} \\
\midrule
\multirow{3}{*}{T1}
& Misplaced Instrument
& \added{When preparing to cut the skin-layer sutures}
& The scissors are not on the instrument table(~\autoref{fig:teaser}\teasericon{3a}); the trainee must go to the supply table(~\autoref{fig:teaser}\teasericon{3b}) to retrieve them before proceeding.
& decision lag (Ch2) \\
& Instrument Drop and Contamination
& \added{When the trainee is about to use forceps to remove the clot}
& The forceps accidentally fall to the floor and are contaminated; the trainee must choose to fetch a sterile replacement from the supply table or switch to manual removal.
& decision lag (Ch2) \\
& Unstable Surgical Lighting
& \added{During wound management}
& The surgical lamp(~\autoref{fig:teaser}\teasericon{1}) suddenly flickers or dims; the trainee may activate an alternative light source or continue operating under restricted visibility.
& cognitive freeze (Ch1), decision lag (Ch2) \\
\midrule
T2 
& Sudden Vitals Deterioration
& \added{Randomly triggered between 1 and 2 minutes into the procedure}
& The monitor(~\autoref{fig:teaser}\teasericon{2}) emits sharp alarms; heart rate increases to 130, SpO$_2$ drops to 80\%; the remaining time on the task countdown is reduced by 2 minutes.
& cognitive freeze (Ch1), decision lag (Ch2) \\
\midrule
T3
& No Improvement After Clot Removal
& \added{After removing the clot}
& The patient's SpO$_2$ remains low and their heart rate remains high. The trainee must independently decide whether to call the anesthesia team for a secondary intervention and prepare intubation equipment.
& decision lag (Ch2), social isolation (Ch3) \\
\bottomrule
\end{tabular}
\end{table*}

\subsection{System Design}

\subsubsection{System Overview}

\par The system is divided into three parts. The trainee first enters the \textbf{Tutorial Stage}, whose goal is to help learners become familiar with the operation procedures and how to interact with the virtual world. Following the tutorial, the trainee completes a \textbf{Preference Questionnaire}, which is used to determine intervention strategies tailored to their personal characteristics. Next, the trainee proceeds to the \textbf{Practice Stage}, where stress-inducing events are introduced while the system continuously monitors physiological signals in real time. When stress levels exceed the baseline threshold, the system automatically triggers and delivers appropriate intervention strategies, guided by the trainee's stated preferences, to support stress regulation. This workflow is enabled by a modular system architecture that integrates scene management, stress perception, and personalized intervention.

\subsubsection{Tutorial Stage and Preference Extraction}

\par The tutorial scene serves as the entry point to the VR training and is primarily designed to familiarize trainees with basic interaction techniques. The scene begins with a guided onboarding process facilitated by a virtual nurse NPC, who introduces essential interaction methods, such as grasping, moving, and manipulating surgical instruments, and provides an overview of the subsequent training task: managing post-thyroidectomy neck hematoma. This stage is intended to lower the trainee's initial cognitive load and promote spatial awareness and operational familiarity through free exploration. After the basic interaction training, the virtual nurse leads the trainee through a standardized walkthrough of the surgical procedure. This walkthrough is conducted in a stress-free and failure-free mode: no stress triggers are activated, and no time constraints are imposed. This design allows trainees to concentrate on procedural learning without external interference.

\par Upon completing the Tutorial Stage, both the trainee and the virtual nurse NPC transition to a Virtual Lounge environment, where the system administers a preference questionnaire entirely within VR to maintain continuity of experience. To ensure scientific rigor and reliability, the questionnaire was developed by integrating three well-established psychological scales: the \textit{Locus of Control (LoC)} scale~\cite{rotter1966generalized}, the \textit{Perceived Need for Structure (PNS)} scale~\cite{neuberg1993personal}, and the \textit{Interpersonal Emotion Regulation Questionnaire (IERQ)}~\cite{hofmann2016interpersonal}. Together, these scales capture key trainee characteristics and preferences in high-pressure environments. The LoC scale identifies whether trainees attribute outcomes internally or externally, informing the balance between autonomous support and direct guidance. The PNS scale captures preferences for clarity and predictability, guiding the level of detail in interventions. The IERQ assesses reliance on social resources for emotion regulation, shaping whether support is emotional or sensory. Responses from this questionnaire drive the system’s adaptive interventions—for instance, trainees with low internal control and high structure needs receive prompt, explicit workflow instructions rather than ambiguous self-regulation cues. The full questionnaire is provided in ~\autoref{appendix:C}.

\subsubsection{Practice Stage: Stress Perception}

\par After completing the questionnaire, the trainee is notified that a new hematoma incident has occurred in the ward and must proceed to handle it. If the questionnaire results indicate that the trainee does not require socio-emotional support, the NPC will not accompany them into the ward, ensuring the trainee manages the situation independently. Once the simulation begins, one of the system's primary functions is to continuously and accurately monitor the trainee's stress state, which serves as the prerequisite for triggering personalized interventions. This section describes how the system employs physiological signal acquisition and analysis to achieve this objective.

\begin{figure}[h]
    \centering
    \includegraphics[width=\columnwidth]{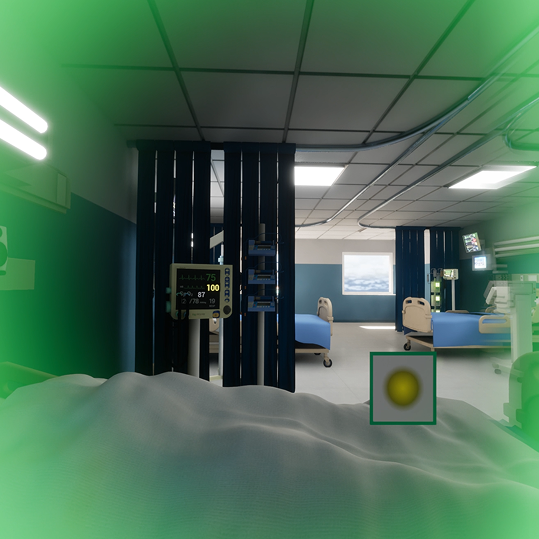}
    \caption{An integrated Self-Regulation Aids interface displaying Breathing Guidance and Stress Feedback.}
    \Description{A first-person view from inside a virtual reality medical simulation showing a hospital room. Two user interface elements for self-regulation are overlaid on the scene: a green, glowing light effect around the edges of the view, which serves as Breathing Guidance, and a small square indicator in the lower-right containing a glowing yellow sphere, which represents Stress Feedback at an 'alert' level.}
    \label{fig:self-segulation aids}
    \vspace{-10pt}
\end{figure}

\par Heart rate (HR), heart rate variability (measured by SDNN), and GSR are reliable physiological indicators for assessing acute stress states. According to ~\citet{kirschbaum1993trier}, under acute stress conditions, individuals typically exhibit a significant increase in heart rate, with the average HR rising by more than 35\%. Based on this evidence and our observations during the system development process, we define a stress state when the heart rate increases by more than 30\%. Furthermore, SDNN has been shown to be strongly associated with acute stress in nursing scenarios. As reported in ~\citet{schneider2017validity}, the mean SDNN decreases by approximately 35\% under acute stress. Therefore, our system sets the SDNN threshold at less than 35\% of the baseline value to determine the presence of stress. In addition, GSR has been widely used in acute stress monitoring ~\cite{turankar2013effects}. In our system, we employ GSR sensors along with a built-in emotional fluctuation detection module to measure variations in skin conductance, thereby enabling more accurate assessments of the user's stress state.

\subsubsection{Practice Stage: Intervention Strategy Design}

\par To ensure high accuracy and sensitivity in stress detection, the system adopts a multi-signal decision-making approach. Specifically, when at least two of the three physiological indicators indicate that the user is under stress, the stress detection module triggers an alert, notifying the user that their stress level is excessively high and intervention may be required. This method helps reduce misjudgments caused by relying on a single indicator and improves overall detection reliability. During practical implementation, participants are required to wear dedicated sensors to ensure data accuracy and minimize artifacts. The electrocardiogram (ECG) sensor is worn on the non-dominant wrist, while the GSR sensor is attached to the ring and little fingers via finger sleeves, preventing interference during handheld operations. After the sensors are in place, the system first performs a 1-minute baseline data acquisition in a resting state, collecting HR and SDNN values to compute the baseline averages. Additionally, to evaluate decision latency, the system monitors the time spent by the user on specific task phases. When the duration of a task exceeds a predefined threshold, the system determines that decision latency has occurred. \added{The decision logic is illustrated in \autoref{appendix:F}.}

\par Through this approach, our stress detection system can monitor participants’ physiological changes in real time and perform precise signal analysis, enabling timely interventions during acute stress states to help users effectively manage and cope with stress.

\par Once the stress perception module detects that a trainee's stress level significantly exceeds their baseline, the system automatically initiates the appropriate intervention process. Unlike traditional single-mode intervention approaches, this system adopts a modular design that tailors intervention strategies based on preference questionnaire results collected after the instructional scenario. These strategies were informed by feedback from senior physicians during focus group discussions and are explicitly aligned with the five proposed design goals: delivering timely (\textbf{DG1}), hierarchical (\textbf{DG2}), personalized (\textbf{DG3}), and minimally disruptive (\textbf{DG4}) support, while promoting the internalization and \replaced{potential transferability}{transfer} of stress-coping skills (\textbf{DG5}).

\par The intervention framework is organized around three complementary dimensions: \emph{Self-Regulation Aids}, \emph{Procedure Guidance}, and \emph{Emotional/Sensory Support}. Depending on the trainee's individual profile, the system dynamically selects and integrates the most suitable combination of interventions, adapting to both the user's state and the situational context.

\begin{figure*}[h]
  \centering
  \vspace{-10mm}
  \includegraphics[width=\textwidth]{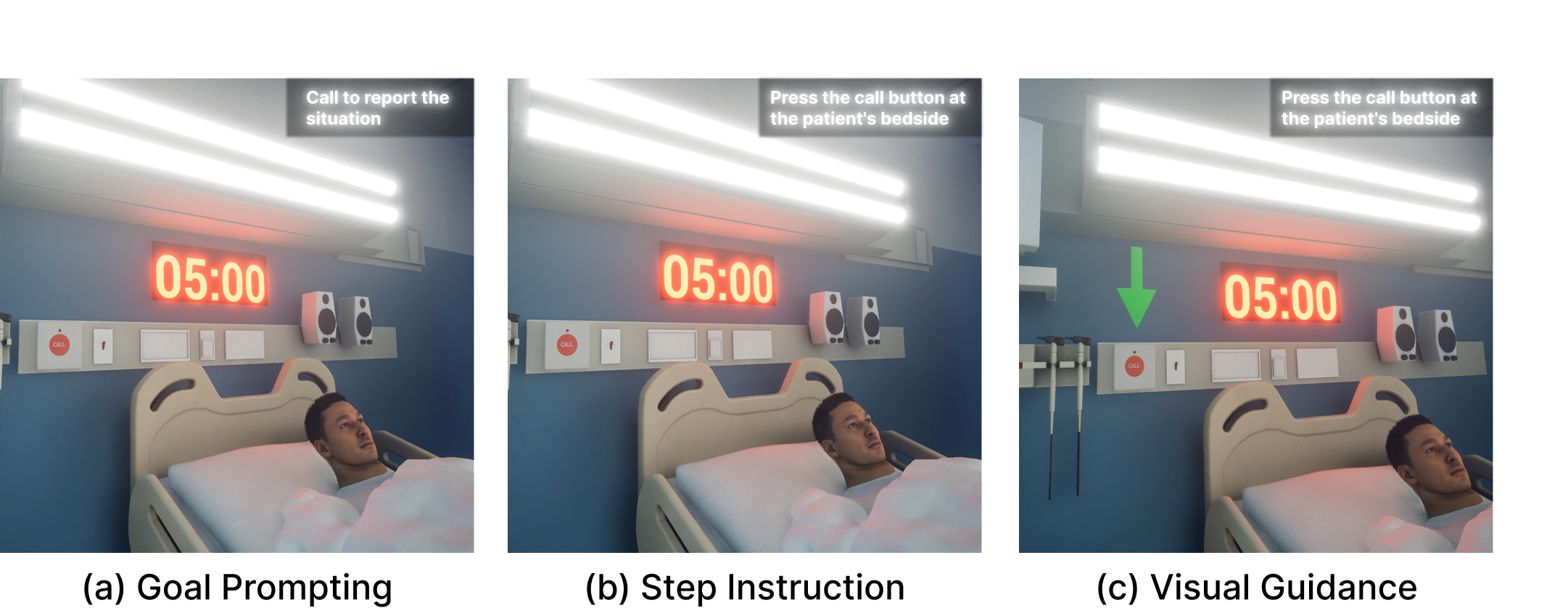}
  \caption{The hierarchical guidance mechanism of Procedure Guidance, escalating from a high-level Goal Prompt (a), to a specific Step Instruction (b), and finally to explicit Visual Guidance (c).}
  \Description{A three-panel diagram illustrates hierarchical guidance system of Procedure Guidance strategy in a simulated hospital room, showing escalating levels of support. Panel (a) Goal Prompting: Displays a high-level instruction in a text box: "Call to report the situation." Panel (b) Step Instruction: Provides a more specific, actionable command: "Press the call button at the patient's bedside." Panel (c) Visual Guidance: Repeats the instruction from panel (b) and adds a large green arrow pointing directly to the call button on the wall to provide explicit visual direction.}
  \label{fig:procedure guidance}
\end{figure*}

\par \capsulebw{Self-Regulation Aids}
This category focuses on helping trainees independently manage stress responses by combining physiological feedback with cognitive guidance, thereby facilitating recovery and supporting rational decision-making. It includes two core intervention modules, as shown in ~\autoref{fig:self-segulation aids}:

\begin{itemize}
    \item \textit{Breathing Guidance.} This module draws on design concepts from visual prototypes VP1 (heart rate waveform) and VP3 (central breathing guidance circle). Iterative testing revealed that VP1's abstract waveform imposed unnecessary cognitive load, while VP3's centrally located prompts interfered with task execution. Therefore, we proposed a compromise solution: when the system detects that the learner's stress exceeds the threshold, breathing guidance light effects will appear at the edges of the virtual field, dynamically expanding and contracting with a smooth, preset rhythm. By situating the prompt outside the core operational area, this design adheres to the low-disruption principle (\textbf{DG4}), implicitly enhancing breathing awareness and supporting effective physiological regulation.
    \item \textit{Stress Feedback.} This module builds on the visual prototype VP2 to heighten trainees' awareness of their stress states with minimal cognitive burden. A small light indicator at the edge of the visual field dynamically shifts color—green (safe), yellow (alert), red (overload)—based on real-time physiological data. This design provides immediate physiological feedback while serving as a controlled VR-based stress inoculation mechanism, enabling trainees to adapt to extreme responses in training. By fostering resilience and operational stability under pressure, it further supports skill internalization and \replaced{fosters the metacognitive awareness necessary for transfer}{transfer} (\textbf{DG5}).
\end{itemize}

\par \capsulebw{Procedure Guidance} This category focuses on reducing trainees' cognitive load during high-pressure operations. To balance effective guidance with learner autonomy, three visual prototypes—task checklist (VP4), guidance arrows (VP5), and HUD prompts (VP6)—are integrated into a hierarchical intelligent guidance module(~\autoref{fig:procedure guidance}). Aligned with the principles of timeliness (\textbf{DG1}) and hierarchical support (\textbf{DG2}), the system dynamically adjusts the level of guidance granularity based on real-time behavioral data.
\begin{itemize}
    \item \textit{Goal Prompting}: At the beginning of a task, the system provides only high-level prompts. A concise checklist is displayed at the periphery of the trainee's visual field, using minimal text (e.g., ``Remove the blood clot'') to clarify objectives while encouraging independent planning and active recall.
    \item \textit{Step Instruction}: If behavioral tracking detects prolonged inactivity or hesitation beyond a preset threshold, the system escalates the intervention. The checklist is updated with more detailed, actionable instructions (e.g., ``Please remove the Steri-strips covering the wound''), thereby reducing cognitive load and facilitating task progression.
    \item \textit{Visual Guidance}: If trainees remain inactive after receiving instructions, the system infers difficulties in target localization or cognitive freezing. The intervention then escalates further, highlighting critical objects or overlaying semi-transparent guidance arrows to indicate the exact target location, ensuring continuity when trainees encounter significant cognitive barriers.
\end{itemize}

\par Intervention thresholds are personalized (\textbf{DG3}) to accommodate individual differences. Trainees with strong independence are given extended wait times, encouraging autonomous problem-solving, whereas those with high dependence or anxiety receive earlier support to mitigate stress overload. This adaptive, layered guidance ensures timely intervention while preserving learner autonomy, effectively balancing structured support with opportunities for self-directed learning.

\par \capsulebw{Emotional/Sensory Support} This strategy addresses stress caused by the combined effects of environmental and emotional factors, helping trainees maintain focus and reduce feelings of isolation during high-pressure operations.
\begin{itemize}
    \item \textit{Auditory Environment Regulation.} Building on visual prototype VP8, this module targets environmental noise, a frequent stressor for novice physicians (as discussed in~\autoref{sec:stressTriggers}). When stress levels exceed baseline, the system selectively suppresses irrelevant noise while preserving task-critical alerts, thereby reducing distractions and supporting sustained concentration.
    \item \textit{Companion-based Emotional Support.} Informed by iterative designs of visual prototype VP9, this module offers emotional support without fostering dependency on the system. Activation is contingent on questionnaire results and applies only to trainees identified as requiring external support in high-pressure contexts. To minimize social uncertainty, the supportive NPC(~\autoref{fig:teaser}\teasericon{5}) is represented by a familiar figure—a nurse who already guides trainees during orientation scenarios—rather than introducing new characters. Once triggered, the NPC delivers non-instructional verbal encouragement (e.g., ``Stay calm, you can handle this'' or ``Focus on the next critical step''), leveraging social presence to reduce feelings of isolation while preserving autonomy in decision-making.
\end{itemize}

\subsection{Implementation Details}
\par The system was implemented in Unreal Engine 5.6.0~\footnote{https://www.unrealengine.com/}. To ensure high-quality visual rendering and stable physical interactions, we developed a software architecture that supports real-time image rendering, precise object manipulation, and seamless data processing. Core interaction logic is managed through VR controllers, which enable operations such as grasping, placing, and moving virtual objects. To enhance realism and immersion, the non-player characters (NPCs), including the patient and the nurse, were created using Unreal Engine's MetaHuman framework~\footnote{https://www.metahuman.com/}.

\par For real-time physiological sensing, the system integrates modules for acquiring and processing GSR and Photoplethysmography (PPG) sensor signals. \added{For PPG signal acquisition, we employed the Cheez.PPG wristband sensor}~\footnote{https://github.com/CheezCheez/CheezPPG/tree/main/examples}\added{, which uses green-light photoelectric detection at a sampling rate of 125 Hz. The raw PPG data undergoes a preprocessing pipeline including moving average filtering for noise smoothing, enabling reliable heart rate and HRV (SDNN) extraction through peak detection algorithms. For GSR measurement, we utilized the Grove GSR Sensor(Seeed Studio, Model 101020052)}~\footnote{https://www.seeedstudio.com/Grove-GSR-sensor-p-1614.html}\added{, operating at 5V with analog voltage output. The Grove SDK}~\footnote{\url{https://wiki.seeedstudio.com/Grove-GSR_Sensor/}} \added{provides built-in emotional fluctuation detection, which outputs signal feedback when emotional changes are detected, allowing the system to respond to users' affective state variations in real time.} These data are processed locally and analyzed alongside interaction data to drive adaptive feedback and, when necessary, activate intervention strategies. To safeguard user privacy, all data remains confined to the training session: they are neither stored locally nor transmitted to external servers.

\section{User Study}

\begin{figure*}[h]
    \centering
    \includegraphics[width=\textwidth]{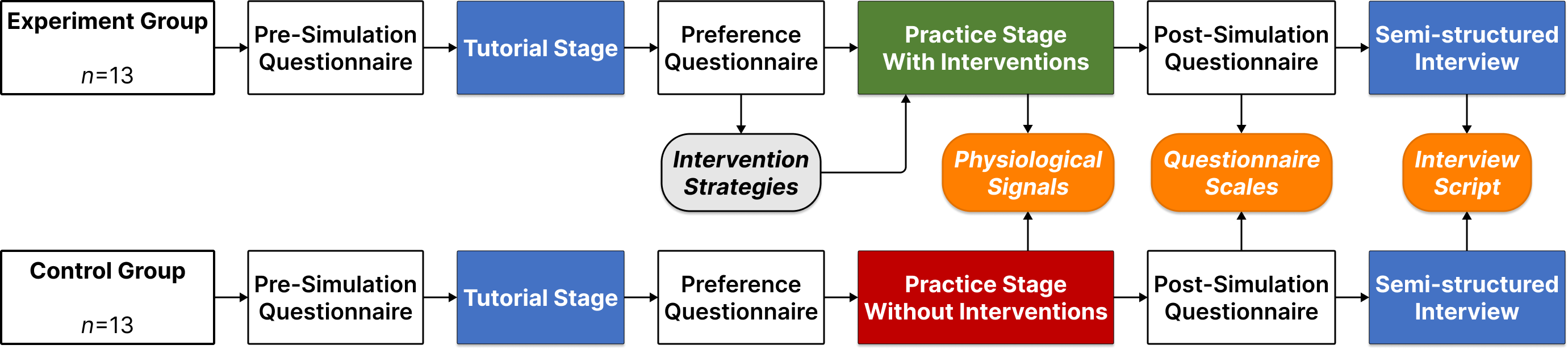}
    \caption{An overview of the two-group between-subject user study. Both groups followed a standardized procedure, with the key difference occurring in the Practice Stage, where the Experimental Group received real-time interventions. The diagram also illustrates the key data collected to evaluate the system's impact on \textbf{Task Performance}, \textbf{Cognitive Recovery} (via physiological signals), and \textbf{Subjective Experience} (via questionnaire scales and interviews).}
    \Description{A flowchart illustrating the procedure of a two-group, between-subject user study. The chart displays two parallel rows representing the Experiment Group (n=13) on top and the Control Group (n=13) at the bottom. Both groups follow the same sequence of stages: Pre-Simulation Questionnaire, Tutorial Stage, Preference Questionnaire, a Practice Stage, Post-Simulation Questionnaire, and a Semi-structured Interview. The key difference is highlighted in the Practice Stage: the Experiment Group's stage is green and labeled "Practice Stage With Interventions," while the Control Group's is red and labeled "Practice Stage Without Interventions." The diagram also indicates that data such as Physiological Signals, Questionnaire Scales, and Interview Scripts were collected to evaluate the system's impact on Task Performance, Cognitive Recovery, and Subjective Experience.}
    \label{fig:userstudy}
\end{figure*}

\par To systematically evaluate the effectiveness of our system, we conducted a user study aimed at addressing \textbf{RQ3}: \textit{How does the system influence trainees' task performance, cognitive recovery, and subjective experience under acute stress?} To this end, we adopted a two-group between-subject design, randomly assigning participants to either the \textbf{experimental group} (receiving real-time intervention) or the \textbf{control group} (no intervention).

\subsection{Participants and Experiment Setting}

\subsubsection{Participants}
\par We recruited 29 participants with medical backgrounds through posters distributed across three local partner hospitals \added{and via snowball sampling. All participants met the ``novice physician'' criteria defined in ~\autoref{sec:Formativestudy}, and none had participated in the prior formative study.} Of these, 26 completed the study \added{($12$ males, $14$ females; age $M=24.92, SD=1.06$)}, while three withdrew due to scheduling conflicts. \replaced{To minimize confounding variables, participants were randomly assigned to either the experimental group ($P1-P13$) or the control group ($P14-P26$) using a block randomization strategy, stratified by department and years of experience. Independent samples t-tests were performed to compare the two groups. No significant differences were found in age($\textit{p} > .05$, $\textit{d}=0.14$) or prior VR experience ($\textit{p} > .05$, $\textit{d}=0.26$).}{All participants were either novice physicians or medical students currently undertaking clinical rotations, aligning closely with the system's target user population. Participants were randomly assigned to the experimental group (\(P1-P13\)) or the control group (\(P14-P26\)).} 

\par Before beginning the study, all participants were briefed on its objectives and procedures and provided written informed consent. Those who completed the study received a \$25 honorarium. The participants’ demographic information can be found in \autoref{appendix:D}.

\subsubsection{Apparatus and Environment}
\par The experiment was conducted in a controlled laboratory setting to ensure consistency and support immersion. Participants remained seated throughout the session and wore an HTC VIVE Focus Vision headset~\footnote{https://www.vive.com/us/product/vive-focus-vision/overview/}, tethered to a high-performance PC equipped with an NVIDIA RTX 4080 Super GPU for stable streaming. Handheld controllers were used for interactions such as grasping and placing objects.

\par To capture physiological states in real time, GSR and PPG sensors were attached to participants' non-dominant hands, recording GSR, HR, and SDNN. These signals were transmitted to the PC for real-time analysis, which enabled adaptive feedback and triggered intervention strategies for the experimental group when stress levels exceeded predefined thresholds. The study took place in an approximately $10\,\mathrm{m}^2$ private and quiet room to minimize external disturbances and foster an immersive experience.

\subsection{Procedure}
\par The experimental procedure followed a standardized protocol to ensure consistency across participants. It comprised four stages: \textit{preparation and pre-test}, \textit{tutorial}, \textit{practice task}, and \textit{post-test with interviews}. The entire study, from arrival to interview completion, lasted approximately \textbf{60 minutes}. ~\autoref{fig:userstudy} shows the overall procedure of the user study.

\subsubsection{Preparation and Pre-test}
\par Upon arrival, participants were briefed on the study and provided written informed consent. They then completed a pre-simulation questionnaire consisting of three components: (a) demographic information, VR usage experience, and medical background; (b) knowledge-based questions on managing postoperative neck hematoma to assess baseline expertise; (c) the \textbf{STAI-S} state anxiety scale to measure immediate anxiety levels prior to the experiment. \added{Statistical analysis of the pre-test data showed no significant differences between the experimental and control groups in terms of baseline domain knowledge scores($\textit{p} > .05$, $\textit{d}=0.13$) or initial state anxiety levels ($\textit{p} > .05$, $\textit{d}=0.14$).}

\subsubsection{Tutorial Stage}
\par Participants were equipped with the VR headset and physiological sensors before entering the Tutorial Stage. After completing the tutorial, they verbally summarized the core procedural steps to confirm comprehension. They then filled out a preference questionnaire within the VR environment. For the experimental group, responses were used to configure personalized intervention strategies, while the control group completed the same questionnaire without activating the intervention module.

\subsubsection{Practice Stage}  
\par In the Practice Stage, participants independently performed a standardized emergency procedure under simulated stress. To ensure \replaced{consistency of experimental conditions}{consistent stress induction}, all stressors are activated, and the Sudden Vitals Deterioration event is triggered exactly one minute after the scenario begins, which means participants only have \textbf{3 minutes} to complete all tasks. Tasks were marked as incomplete if participants failed to complete all required steps within the allotted time. Physiological signals of participants in both groups will be recorded throughout the entire process, with the experimental group receiving system intervention when high-pressure states are detected, while the control group will not receive system intervention.

\subsubsection{Post-test and Interview}  
\par Following the practice task, participants removed the equipment and completed a post-simulation questionnaire comprising three standardized scales:  
\textbf{NASA-TLX}~\cite{hart1988development}, \textbf{SUS}~\cite{brooke1996sus}, and \textbf{IPQ}~\cite{schubert2001experience}. A semi-structured interview of approximately 15 minutes was then conducted to gather participants' subjective experiences, perceived skill acquisition, and feedback on the intervention strategies.

\subsection{Measures and Data Analysis}

To investigate \textbf{RQ3}, we collected and analyzed experimental data. First, we focused on participants' \textbf{task performance} to evaluate the impact of the system's real-time interventions on operational efficiency and accuracy. Specifically, we examined task completion, operation duration, and critical error rates: whether participants completed all core operations within the allotted time, the total duration from task initiation to the final step, and the occurrence of procedural errors during the operation (e.g., improper use of instruments, reversed or omitted critical steps). These indicators collectively reflect the direct influence of system interventions on operational quality. Regarding \textbf{cognitive recovery}, we calculated the \textit{average recovery time} based on physiological signals (HR, SDNN, and GSR) to assess how quickly participants returned from a high-stress state to baseline levels during the task, thereby evaluating the effectiveness of intervention strategies. Furthermore, we conducted a comprehensive analysis of participants' \textbf{subjective experience} by integrating post-task questionnaires and qualitative interview data. Using the NASA-TLX, SUS, and IPQ scales, we quantified participants' cognitive load, system usability, and sense of immersion, respectively. Meanwhile, we transcribed semi-structured interview recordings and applied thematic analysis to extract key insights regarding the effectiveness of intervention strategies, as well as the strengths and limitations of the system.

\section{Results}

\par To comprehensively evaluate the intervention's effects, we analyzed the experimental data across three levels: objective \textbf{Task Performance}, \textbf{Cognitive Recovery} from high stress, and participants' \textbf{Subjective Experience}. The comparative results for each are presented in the following subsections.

\subsection{Task Performance}

\begin{table}[h]
    \centering
    \caption{Comparison of Task Performance Between Experimental and Control Groups.}
    \label{tab:task_performance}
    \small 
    \begin{tabular}{@{}lccc@{}}
    \toprule
    \textbf{Metric} & \textbf{Exp. Group} & \textbf{Ctrl. Group} & \textbf{$p$-value} \\ \midrule
    Task completion rate & 69.23\% & 15.38\% & -- \\
    Mean operation duration (s) & 164.17 & 207.31 & $< .001$ \\
    Mean no. of critical errors & 0.308 & 0.769 & $>.05$ \\ \bottomrule
    \end{tabular}
\end{table}

\par We evaluated task performance using key metrics, including completion rate, operation duration, and critical errors, as summarized in ~\autoref{tab:task_performance}. The results indicate that participants in the experimental group, who received real-time adaptive interventions, outperformed those in the control group across multiple dimensions. Specifically, 69.23\% of participants (9 out of 13) in the experimental group completed all critical operations within three minutes, compared to only 15.38\% (2 out of 13) in the control group. In terms of operational efficiency, participants in the experimental group who completed the task required significantly less time (\textit{M} = 164.17\added{s}, \textit{SD} = 25.54\added{s}) than their counterparts in the control group (\textit{M} = 207.31\added{s}, \textit{SD} = 30.85\added{s}), with an independent-samples t-test confirming this difference as statistically significant (\textit{p} < .001\added{, \textit{d} = 1.53}). For critical errors, the experimental group exhibited a lower average error rate (\textit{M} = 0.308) compared to the control group (\textit{M} = 0.769), although this difference did not reach statistical significance (\textit{p} > .05\added{, \textit{d} = 0.52}).

\subsection{Cognitive Recovery}

\begin{figure*}[t]
    \centering
    \begin{subfigure}{\textwidth}
        \centering
        \includegraphics[width=\textwidth]{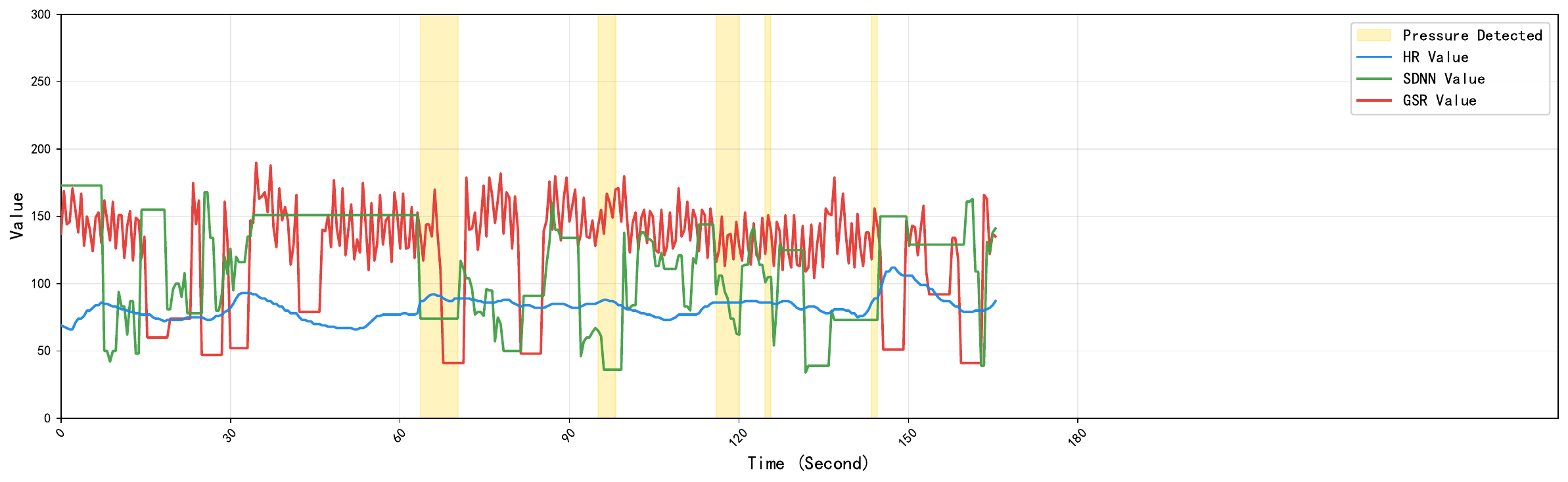}
        \caption{Recording of P6.}
        \label{fig:subfig1}
    \end{subfigure}
    \begin{subfigure}{\textwidth}
        \centering
        \includegraphics[width=\textwidth]{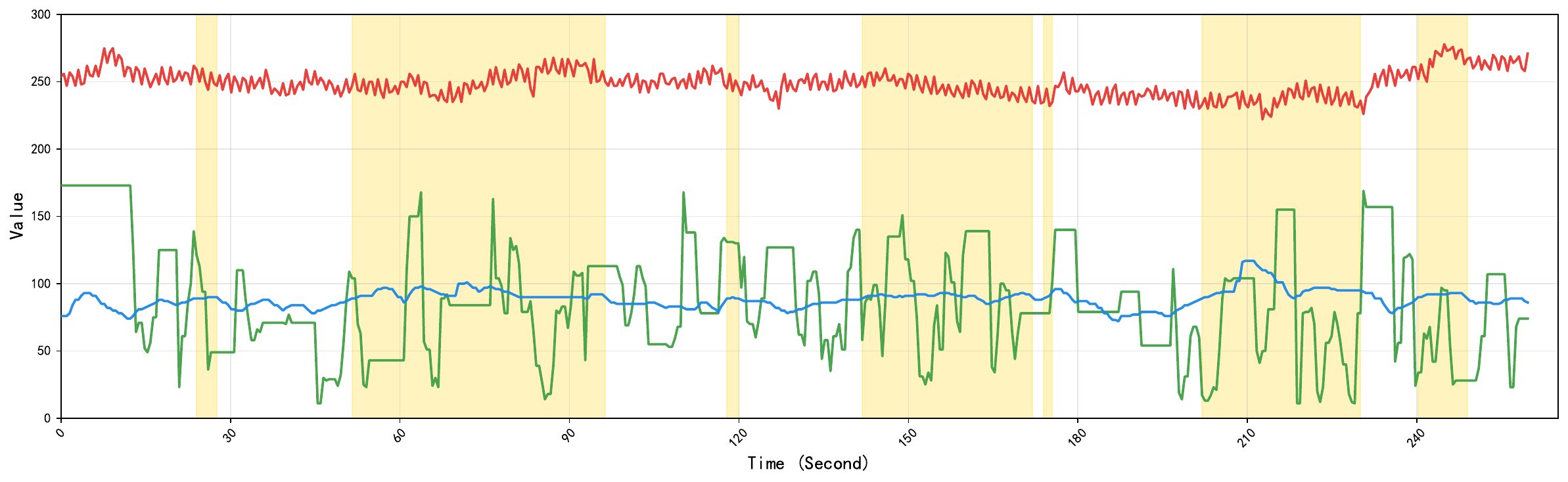}
        \caption{Recording of P25.}
        \label{fig:subfig2}
    \end{subfigure}
    \caption{\replaced{Physiological}{Stress level} recordings for an experimental-group participant (P6) and a control-group participant (P25) during the Practice Stage. P6 completed the task in 163 seconds, and P25 completed the task in 255 seconds. \replaced{The yellow area indicates that the system has detected that the participant is in a high-stress state, and its length reflects the participant's speed of cognitive recovery}{The time segments marked in yellow indicate periods when the system detected that the participant was in a high-stress state}.}
    \Description{A figure with two line graphs comparing the physiological stress recordings of an experimental participant (P6) and a control participant (P25). Both graphs plot Heart Rate (HR), SDNN, and Galvanic Skin Response (GSR) against time. Yellow shaded areas indicate periods of high stress detected by the system. Graph (a) shows participant P6 (experimental group) completing a task in 163 seconds. It displays several brief, distinct periods of high stress. Graph (b) shows participant P25 (control group) completing the same task in 255 seconds. It features longer, more sustained periods of high stress. The comparison visually demonstrates that the experimental participant recovered from stress more quickly than the control participant.}
    \label{fig:combined}
\end{figure*}

\par To assess the impact of system interventions on trainees' cognitive regulation, we measured the average recovery time required for participants to return to baseline following periods of high stress. Participants in the experimental group, who received real-time interventions, demonstrated a significantly shorter mean recovery time (5.09 s, \textit{SD} = 2.36 s) compared to the control group (11.19 s, \textit{SD} = 2.84 s). An independent-samples t-test confirmed that this difference was statistically significant (\textit{p} < .001\added{, \textit{d} = 2.33}), indicating that the intervention effectively accelerated cognitive recovery. ~\autoref{fig:combined} illustrates stress-level trajectories during the \textit{Practice Stage} for a representative experimental participant (P6) and a control participant (P25), highlighting the faster recovery observed in the experimental group after peak stress episodes.

\subsection{Subjective Experience}

\subsubsection{Quantitative Findings}

\begin{figure*}[h]
    \centering
    \includegraphics[width=\textwidth]{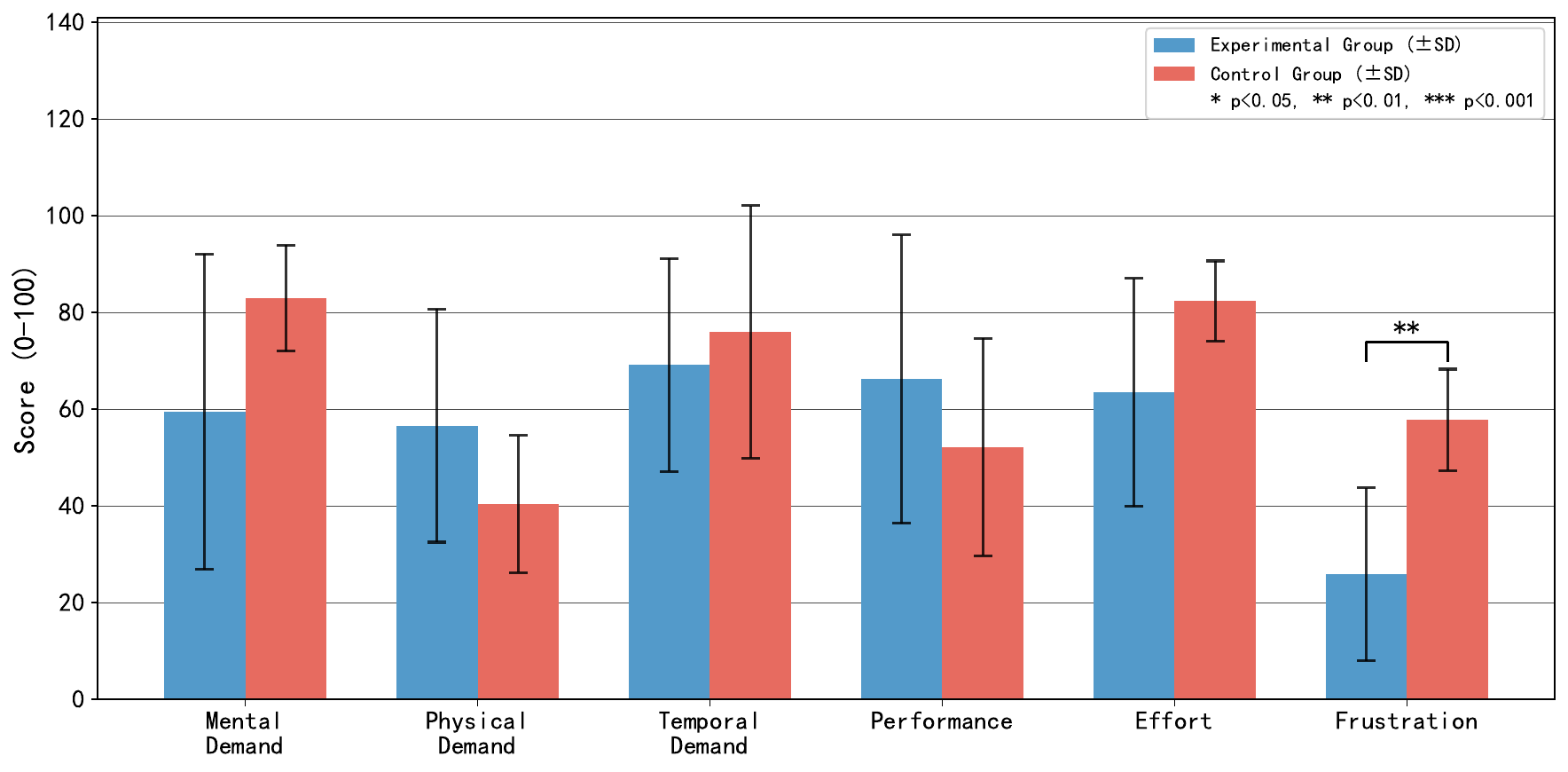}
    \caption{Results for the NASA-TLX portion of the post-simulation questionnaire. Participants' responses were integers from 0 to 100.}
    \Description{A clustered bar chart showing the results of the NASA-TLX questionnaire, comparing the mean scores (from 0 to 100) of an experimental group and a control group across six dimensions. Error bars represent standard deviation. The data for each dimension is as follows: Mental Demand: The experimental group's score (approximately 60) is lower than the control group's (approximately 82). Physical Demand: The experimental group (approximately 57) scores higher than the control group (approximately 40). Temporal Demand: The experimental group (approximately 69) scores slightly lower than the control group (approximately 75). Performance: The experimental group (approximately 66) scores higher than the control group (approximately 52). Effort: The experimental group (approximately 63) scores lower than the control group (approximately 82). Frustration: The experimental group shows a significantly lower score (mean = 25.85) compared to the control group (mean = 57.77).}
    \label{fig:nasatlx}
\end{figure*}

\added{We performed independent-samples \textit{t}-tests on the post-simulation questionnaires to examine subjective experiences across three dimensions: cognitive load, system usability, and immersion. }

\added{\textit{Cognitive Load (NASA-TLX):} The experimental group reported lower overall cognitive load (\textit{M} = 51.38, \textit{SD} = 16.46) compared to the control group (\textit{M} = 64.62, \textit{SD} = 12.83). Although the overall difference did not reach statistical significance (\textit{p} > .05), the effect size was large (\textit{d} = 0.86). Notably, on the specific subscale of ``frustration,'' the experimental group (\textit{M} = 25.85) scored significantly lower than the control group (\textit{M} = 57.77, \textit{p} < .01, \textit{d} = 2.04), with similar trends observed for ``reduced effort'' and ``mental demand'' (see ~\autoref{fig:nasatlx}). }

\added{\textit{System Usability (SUS):} No significant differences were found between the experimental group (\textit{M} = 72.81, \textit{SD} = 9.01) and the control group (\textit{M} = 74.50, \textit{SD} = 14.94; \textit{p} > .05, \textit{d} = 0.15), indicating consistent usability across conditions.}

\added{\textit{Immersion (IPQ):} Similarly, no significant differences were found in overall immersion measurements (\textit{p} > .05, \textit{d} = 0.45). However, regarding the distinction between virtuality and reality, participants in the control group were significantly more likely to report difficulty distinguishing between the two (\textit{p} < .05, \textit{d} = 1.42). Detailed results are provided in \autoref{appendix:E}.}

\deleted{We performed an independent-samples \textit{t}-test on the post-simulation questionnaire data to examine differences in subjective experience between the experimental and control groups. The overall cognitive load reported by the experimental group (\textit{M} = 51.38, \textit{SD} = 16.46) was lower than that of the control group (\textit{M} = 64.62, \textit{SD} = 12.83), although this difference did not reach statistical significance (\textit{p} > .05). Notably, in the ``frustration'' dimension, the experimental group (\textit{M} = 25.85) scored significantly lower than the control group (\textit{M} = 57.77, \textit{p} < .01), with trends also observed toward ``reduced effort'' and ``mental demand'' (see \mbox{~\autoref{fig:nasatlx}}). No significant differences were found in system usability between the experimental group (\textit{M} = 72.81, \textit{SD} = 9.01) and the control group (\textit{M} = 74.50, \textit{SD} = 14.94; \textit{p} > .05), nor in immersion measurements (\textit{p} > .05). However, participants in the control group were significantly more likely to report difficulty distinguishing between virtuality and reality (\textit{p} < .05). Detailed post-simulation questionnaire results are provided in \mbox{\autoref{appendix:E}}.}

\subsubsection{Qualitative Findings}
\par Thematic analysis of the interview transcripts identified two themes characterizing participants' experiences with the intervention.

\par \textit{Theme 1: Diverse Intervention Strategies Support Comprehensive Intervention.} Participants in the experimental group reported that the intervention strategies offered a variety of tools to manage emotions and cognitive load effectively in high-pressure task contexts. Some participants monitored the light spot in the bottom-right corner of the interface, which reflected real-time stress levels, using it as a a cue for self-regulation. As P2 noted: ``\textit{I would glance at it from time to time, and if I saw the light turn red, I reminded myself to calm down.}'' This immediate feedback enabled active emotional regulation. Others relied primarily on the guided breathing feature; P12 explained, ``\textit{When the breathing wave appeared, I realized that my breathing had already sped up.}'' In addition, the system's layered guidance during moments of task stagnation was perceived as effective support. P7 commented, ``\textit{When I was unsure about the next step, my first reaction was to check if there were any prompts.}'' Several participants also highlighted the positive impact of encouraging messages from the nurse NPC. P8 emphasized, ``\textit{in emergency situations, having someone help clarify your thoughts and provide comfort is essential,}'' while P5 added, ``\textit{Hearing some reassuring words mixed with the noisy environment made me feel safer.}''

\par \textit{Theme 2: \replaced{Perceived Transferability}{Transferability} of Stress Regulation Skills.} Participants in the experimental group reported that the intervention not only enhanced their ability to manage postoperative thyroid hematomas but also \replaced{provided strategies they perceived as applicable to broader}{equipped them with transferable strategies for} stress management and crisis response. P3 explained, ``\textit{This improved my ability to respond in this specific context and will also help in other similar medical crises.}'' When asked about the applicability of decompression techniques, P1 confirmed that \replaced{they felt the techniques}{they} ``\textit{can be applied elsewhere''} and elaborated that in different emergency contexts, ``\textit{this method can be used to adjust and adapt.}'' P9 echoed this view, highlighting the broader value of stress reduction skills: ``\textit{Although this clinical scenario is relatively rare, what matters is that you need to maintain yourself under high pressure. The system-provided breathing guidance is very helpful; it allows me to stay effectively focused.}''
In contrast, participants in the control group demonstrated a narrower perception of transferability, often treating the acquired knowledge as a solution tailored to a single problem rather than a framework for broader crisis management. As P24 remarked, ``\textit{After this experience, I can handle this problem, but if I face a new situation, I still wouldn't know what to do.}''

\section{Discussion}

\par \added{This study aims to address a critical gap in real-time cognitive support within VR-based stress inoculation training. In response to \textbf{RQ1}, our formative study identified that novice physicians, in managing acute cognitive overload, require a personalized combination of \textit{self-regulation aids}, \textit{procedure guidance}, and \textit{emotional/sensory support}. These needs were distilled into five design goals (\textbf{DG1}–\textbf{DG5}), prioritizing a balance between immediate assistance and learner autonomy. Regarding \textbf{RQ2}, we demonstrated that the JITAI framework effectively operationalizes these goals. By integrating physiological sensing with intervention mechanisms, our system adapts the traditional SIT paradigm to deliver context-aware support, seamlessly integrated into the surgical workflow. Finally, in response to \textbf{RQ3}, our empirical evaluation revealed that this adaptive approach significantly accelerates cognitive recovery and reduces subjective frustration, enabling trainees to maintain self-control even under high-pressure conditions.}

\subsection{Key Findings}

\begin{figure*}[t]
    \centering
    \includegraphics[width=\textwidth]{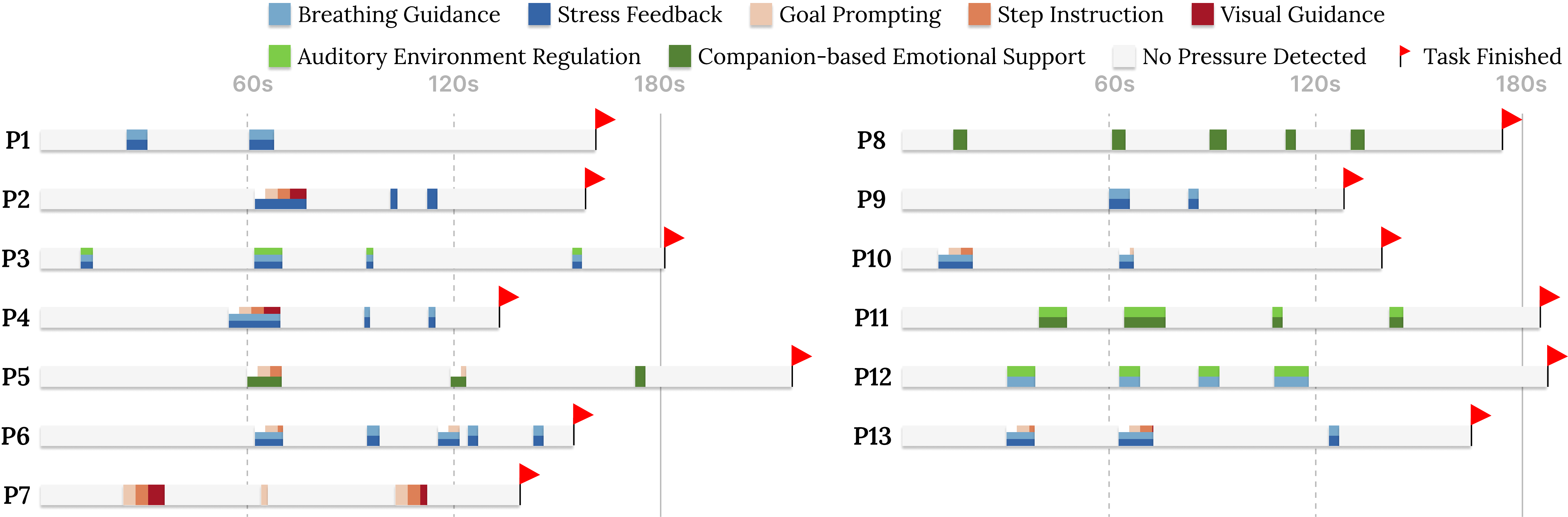}
    \caption{\added{Visualization of the intervention timeline for the experimental group ($N=13$) during the Practice Stage shows that all participants triggered an intervention at approximately 60 seconds. This timing corresponds to the Sudden Vitals Deterioration event, which was programmed to occur one minute into the experiment and was sufficiently intense to elicit a stress response in all participants. Once the system detected a stress response, it delivered interventions based on the strategy combination determined from each participant's preference extraction questionnaire. For example, P1 scored ``Internal'' on the LoC section of the questionnaire; the IERQ responses indicated no objection to breathing guidance or stress feedback, but a preference against noise reduction or external emotional support; and her PNS score was below 15. Consequently, the system provided P1 with an intervention consisting solely of breathing guidance and stress feedback.}}
    \Description{A horizontal visualization split into two columns (Left: P1-P7, Right: P8-P13). The horizontal axis represents the timeline of the 3-minute practice task, marked at 60s, 120s, and 180s. Each row represents a participant's session. Light grey bars indicate periods of "No Pressure Detected". Colored blocks indicate active interventions triggered by the system. A legend at the top distinguishes seven intervention types: Breathing Guidance (light blue), Stress Feedback (dark blue), Goal Prompting (very light red), Step Instruction (light red), Visual Guidance (red), Auditory Regulation (light green), and Companion Support (dark green). A red flag icon marks the "Task Finished" point for each user.}
    \label{fig:InterventionTimeline}
\end{figure*}

\subsubsection{\replaced{Trait-Strategy Alignment: Personalizing Interventions via Psychological Profiling}{Emotion Regulation Under Cognitive Load: The Role of Interventions in VR-SIT}}

\par \added{Our evaluation demonstrates that the efficacy of VR-SIT is not driven by a single dominant intervention component, but by the alignment of intervention strategies with learners' psychological traits. As shown in Fig.~\ref{fig:InterventionTimeline}, intervention patterns varied significantly across participants, reflecting their pre-assessed preferences. For example, P1, who exhibited a high Internal Locus of Control in the preference extraction questionnaire, engaged primarily with non-intrusive support, such as \textit{Breathing Guidance} and \textit{Stress Feedback}. This minimal intervention pattern aligns with a preference for autonomy and independent problem-solving. In contrast, P11 and P12 required frequent, multimodal interventions, combining \textit{Auditory Environment Regulation} and \textit{Companion-based Emotional Support}, indicating a greater need for external structure and social regulation. Additionally, P7 primarily relied on \textit{Procedure Guidance}, reflecting a profile that values procedural clarity over emotional buffering.}

\par \added{This heterogeneity highlights a methodological insight: evaluating the contribution of individual components in isolation yields limited value in an adaptive system. Since different users rely on distinct modules to achieve the same goal of regulation, the system's efficacy stems from its holistic adaptability rather than the standalone power of any single feature. By aligning the support mechanisms with each trainee's inherent coping style, the system allowed participants to maintain a sense of control and avoid decision paralysis. These findings resonate with Gross's Process Model of Emotion Regulation discussed in section ~\ref{sec:2.3.2}: users with high internal control preferred antecedent-focused strategies (e.g., breathing cues) to preemptively manage arousal, while those needing external structure benefited from response-focused support (e.g., companion guidance) during critical moments. This validates our methodological choice to ground JITAI designs in established psychological constructs, ensuring that stress management tools complement, rather than conflict with, the user's natural coping instincts. Nevertheless, we acknowledge that future dismantling studies quantifying individual component contributions could offer granular insights to further optimize the efficiency of such intervention designs.}

\deleted{Our findings suggest that the performance improvements observed in the intervention group were not a result of reduced task difficulty, but rather of reshaped psychological experiences. Although overall cognitive load ratings did not differ significantly between groups, participants who received interventions reported markedly lower frustration levels. By offering multi-layered situational support, the system enabled participants to maintain a sense of control and avoid becoming overwhelmed under pressure. This allowed them to engage more effectively with training resources, whereas control group participants were more prone to decision paralysis. These results indicate that VR-SIT design should move beyond merely minimizing objective cognitive load and instead prioritize regulating negative emotional states, thereby fostering sustained engagement and psychological resilience in high-stress contexts.}

\subsubsection{JITAI Framework: Enhancing Learning Efficiency Through Selective Intervention}  
\par The psychological benefits were achieved through the targeted application of a JITAI framework. Rather than offering continuous assistance, JITAI leverages selective engagement, delivering minimal yet \replaced{targeted}{precise} support only when elevated stress levels are detected. This approach significantly improved efficiency: trainees in the intervention group required nearly half the time to regain control compared to the control group. By reducing unnecessary disruptions and alleviating persistent cognitive demands, JITAI optimized high-efficiency learning windows. During these windows, participants actively processed information and consolidated skills, rather than being hindered by unmanaged stress. These findings underscore a critical implication for VR-SIT design: interventions should not be ever-present, but instead strategically emerge at moments of greatest learner need.

\subsubsection{Controlled Presence: Rethinking Immersion in VR-Based Stress Training}  
\par Our findings reveal a nonlinear relationship between immersion and training effectiveness, challenging the assumption that greater immersion automatically yields better outcomes. Participants in the control group reported stronger confusion between virtual and physical contexts, often considered a marker of heightened immersion, yet their task performance was notably poorer than that of the intervention group. We argue that the lack of timely support \replaced{was associated with cognitive overload, which may undermine metacognitive regulation}{led to cognitive overload, undermining metacognitive regulation}. In contrast, the intervention group sustained a state of controlled presence: sufficiently immersed to evoke realistic stress, yet able to analyze contexts, execute procedures, and learn from experience. These results call for a more nuanced approach to VR training design, highlighting the need to balance immersive realism with metacognitive manageability.  

\subsection{\added{Unique Affordances of VR-based JITAI in Medical Training}}

\par \added{While JITAI has been extensively studied in mobile health and wearable contexts for chronic stress management, our study emphasizes the distinct advantages of implementing this framework within VR for high-stakes medical training. The key distinction lies in the \textit{granularity of context awareness} and the \textit{modality of intervention}. Traditional mobile-based JITAI typically relies on generalized physiological triggers or self-reports, delivering interventions through explicit notifications (e.g., pop-ups) that require a shift in user attention. Such mechanisms are fundamentally incompatible with the ``hands-busy, eyes-busy'' nature of surgical emergencies, where task continuity is critical.}

\par \added{In contrast, VR offers a closed-loop ecosystem where the system acts both as the stressor generator and the intervention provider. This aligns with the vision of Cognition-Aware Computing reviewed in section ~\ref{sec:2.3.1}, where systems must sense and adapt to cognitive states in real-time. VR enables the orchestration of acute stress events tailored to specific learning objectives. This fully controllable environment captures not only physiological changes but also detailed behavioral contexts, allowing for the deployment of \textit{implicit, ambient interventions}. Examples from our system include peripheral breathing cues and noise filtering, which integrate seamlessly into the clinical workflow without disrupting immersion or requiring cognitive disengagement.}

\par \added{Moreover, this integration provides unique value to medical education by bridging the gap between theoretical stress management strategies and real-world practice. In traditional medical simulations, stress management is typically addressed in post-hoc debriefings. However, VR-JITAI establishes a ``safety net'' within the simulation itself. It enables novice physicians to experience the physiological onset of acute stress while receiving real-time support to facilitate recovery, effectively training the metacognitive loop of stress regulation in a way that static platforms cannot replicate. This suggests that VR is not merely a visualization tool, but a uniquely effective platform for operationalizing the \textit{skill acquisition and rehearsal} phase of SIT.}

\subsection{Fostering Transferable Skills}
\subsubsection{Beyond Procedural Mastery: Scaffolding Metacognitive Skills for Crisis Management}

\par A key contribution of this study lies in \replaced{highlighting}{demonstrating} the potential of JITAI to foster transferable metacognitive abilities in addition to improving immediate task performance. Evidence from semi-structured interviews revealed that participants in the intervention group not only mastered operational procedures but also \replaced{began to develop a generalized framework}{acquired a generalized framework} for crisis management. They \replaced{reported perceiving}{perceived} the system-modeled stress regulation strategies as broadly applicable to diverse tasks and contexts. In contrast, control group participants tended to focus narrowly on task-specific steps, lacking preparedness for novel scenarios.

\par The JITAI framework can be conceptualized as a dynamic scaffold operating within learners' Zone of Proximal Development (ZPD)~\cite{vygotsky1978mind}. When elevated stress pushes tasks beyond trainees' independent capabilities, the system provides minimal necessary support to return tasks within their manageable range. Beyond facilitating task completion, this approach demonstrates problem-solving processes under pressure, promoting metacognitive awareness. Following the principle of gradual withdrawal~\cite{wood1976role}, intervention support is progressively reduced across repeated training, enabling the internalization and transfer of skills.

\subsubsection{Implications for Medical Education: Toward a Self-Regulated Learning Paradigm}
\par These findings underscore the potential of JITAI-informed VR-SIT to fulfill the core goal of stress inoculation training: equipping individuals with flexible strategies to manage diverse stressors and enhance adaptive capacity. In the context of medical education, this suggests a paradigm shift from procedural compliance toward fostering self-regulatory expertise. The ultimate objective is to prepare professionals capable of maintaining cognitive clarity and making decisive judgments under high-risk, unpredictable conditions. By accelerating cognitive recovery, reducing frustration, and promoting the acquisition of transferable metacognitive strategies, this study outlines a pathway for transforming medical training from mere procedural proficiency into the cultivation of highly adaptive, self-regulated practitioners.

\subsection{Limitation and Future Work}
\par While this study offers preliminary empirical evidence for implementing real-time interventions in VR-SIT, several limitations remain, pointing toward directions for future research.

\subsubsection{\replaced{Technical Implementation and Measurement Constraints}{Interaction Mode}}
\par \replaced{Our system currently employs a wired configuration to ensure stable transmission between the VR headset and physiological sensors. While this setup minimized data loss, it restricted participants' mobility, enforcing a seated operation that diverges from the dynamic movement required in real-world scenarios. Furthermore, despite the stability of wired connections, physiological signals remain susceptible to motion artifacts. Although we applied filtering algorithms, the hand movements inherent in surgical tasks may still introduce measurement noise, potentially affecting the granularity of stress detection.}{To ensure stable data transmission between the VR headset and physiological sensors, we adopted a wired configuration. Although this reduced the risk of data loss, it restricted participants' mobility, resulting in seated operations that differ from real-world scenarios where physicians must move freely around the patient's bed.} Moreover, the use of handheld controllers for fine manipulations may have introduced additional learning costs compared with emerging hand-tracking technologies, thereby compromising interaction naturalness.

\added{Additionally, the current intervention triggering mechanism relies on fixed physiological thresholds. This rule-based approach may lack the nuance required for highly complex clinical scenarios. Future iterations should leverage the full spectrum of VR telemetry—integrating eye-tracking metrics and fine-grained interaction logs—to construct robust, multi-modal predictive models. This would enable interventions that are contextually predictive and precise rather than merely reactive, moving towards truly timely support.}

\subsubsection{\replaced{Experiential Limitations and Simulation Fidelity}{Environmental Realism}}
\par Despite VR's immersive nature, simulated environments cannot fully replicate the stressors and coping behaviors of real clinical settings. For instance, one participant (P17) reported that when hearing the monitor alarm, his instinctive reaction was to ``\textit{walk up to the monitor to silence the alarm}'', only to realize that the system did not support this action. Such mismatches may disrupt presence and highlight the need for designs that better align operational freedom with realistic behavioral expectations. \added{Moreover, prolonged exposure to VR environments poses the risk of cybersickness, which can confound physiological stress readings and limit training duration. Although no participants withdrew due to discomfort, we did not quantitatively assess cybersickness symptoms using standardized tools like the Simulator Sickness Questionnaire (SSQ). Future studies should rigorously monitor this variable to ensure that physiological responses are attributed to clinical stress rather than discomfort.}

\subsubsection{External Validity and Generalizability}
\par The external validity of this study is limited in two respects. First, the study focused on a single clinical scenario, primarily assessing individual decision-making and procedural skills, while real-world emergencies rely heavily on team collaboration, communication, and task distribution, which are key sources of clinical stress and coping strategies absent from our simulation. Second, we assessed only the immediate effects of interventions in a controlled environment without evaluating long-term skill transfer to clinical practice. \added{Consequently, findings regarding the transferability of stress coping skills rely primarily on participant self-reports and qualitative feedback. While these insights suggest a potential for skill retention, they do not constitute objective evidence of clinical transfer.} Future research should conduct longitudinal follow-ups and extend the framework to team-based, high-risk scenarios to better assess generalizability and real-world impact.

\section{Conclusion}
\par This study addresses the acute cognitive overload challenges junior doctors face in surgical emergencies by introducing a VR-SIT system grounded in the JITAI framework. Through a user study with 26 participants, we demonstrated that real-time adaptive interventions can enhance task performance under stress, accelerate cognitive recovery, and improve subjective experience. Our findings contribute a concrete design paradigm for next-generation medical training systems, shifting from stress-focused simulation to dynamic, personalized cognitive assistance. While constrained interaction modalities and simplified scenarios limit the current implementation, this work lays the foundation for future research on long-term training effects, broader high-risk medical contexts, and integration with advanced biosensors. Ultimately, our study suggests a path toward intelligent training environments that support clinicians in maintaining cognitive clarity and decision-making accuracy at critical moments.

\begin{acks}
We gratefully acknowledge the anonymous reviewers for their insightful feedback. This research was supported by the National Natural Science Foundation of China (No. 62372298), the Shanghai Engineering Research Center of Intelligent Vision and Imaging, the Shanghai Frontiers Science Center of Human-centered Artificial Intelligence (ShangHAI), and the MoE Key Laboratory of Intelligent Perception and Human-Machine Collaboration (KLIP-HuMaCo).
\end{acks}

\bibliographystyle{ACM-Reference-Format}
\bibliography{AAReferences}

\appendix
\clearpage
\section{Visual Prototypes}\label{appendix:B}
\begin{table}[h]
    \centering
    \begin{tabular}{cp{3cm}p{13cm}}
    \hline
    \textbf{ID} & \textbf{Name} & \textbf{Description} \\
    \hline
    VP1 & Heart Rate Waveform & A rhythmic waveform line emerges at the periphery of the visual field, resembling a sinusoidal curve. This dynamic line continuously flows, where the frequency of its peaks and troughs represents the user's current heart rate status: as the heartbeat accelerates, the waveform becomes denser and oscillates more rapidly, evoking a sense of tension or anxiety; when the heart rate stabilizes, the waveform gradually stretches and slows down, providing a visual cue of calmness and relaxation. \\
    \hline
    VP2 & Stress Indicator & A luminous dot appears at the periphery of the visual field to indicate stress levels, using a green-yellow-red color coding scheme. Simultaneously, the overall brightness of the visual field progressively dims with increasing stress, particularly around the edges, simulating the sensation of visual narrowing that occurs during heightened breathing or near-fainting episodes. \\
    \hline
    VP3 & Central Breathing Guidance Circle & A softly pulsating circle emerges at the center of the user’s visual field, expanding and contracting slowly in a rhythmic manner. This guides the user to perform “box breathing” and is automatically triggered when abnormal heart rate or stress levels are detected. \\
    \hline
    VP4 & Task Checklist & A task checklist appears on the right side of the screen, listing the critical steps the system expects the user to complete. Each completed step is automatically checked off. If the system detects prolonged stagnation at any particular step, a visual prompt is displayed. \\
    \hline
    VP5 & Guidance Arrows & Semi-transparent guiding arrows are generated to point toward key objects or instruments required for the current task, with the option for the user to enable or disable this feature. \\
    \hline
    VP6 & HUD prompts & The current task steps are persistently displayed in the upper HUD (Head-Up Display) as clear textual instructions, indicating the next required action. Users can choose whether to enable this function. \\
    \hline
    VP7 & HUD Psycho-physiological Encouragement & Through the HUD, the system provides concise textual prompts reflecting the user’s current stress level, along with positive emotional guidance. For example: “Low stress detected, keep it up!” or “Heart rate increasing, try slowing your breath.” \\
    \hline
    VP8 & Noise Filtering & When elevated stress levels are detected, the system automatically filters out persistent environmental noise sources, such as background conversations, phone ringing, or non-essential instrument sounds, thereby reducing distractions. \\
    \hline
    VP9 & Virtual Companion Support & A non-player character (NPC) nurse appears in the scene when excessive stress is detected. The NPC provides voice-based emotional encouragement in a gentle tone to help the user regain composure. \\
    \hline
    \end{tabular}
\end{table}

\clearpage

\section{Preference Extration Questionnaire}\label{appendix:C}

A hybrid approach combining three questionnaires – LoC (Locus of Control), PNS (Perceived Need for Structure), and IERQ (Interpersonal Emotion Regulation Questionnaire) – is adopted to assess user preferences for different types of intervention strategies in high-pressure task scenarios.

\begin{itemize}
    \item \textbf{LoC (Locus of Control):} Differentiates whether participants tend towards ``self-regulation'' or ``reliance on external control''.
    \item \textbf{PNS (Perceived Need for Structure):} Assesses the strength of the need for ``operational process guidance'' in high-pressure tasks.
    \item \textbf{IERQ (Interpersonal Emotion Regulation Questionnaire):} Assesses participants' need for ``emotional/sensory support'' in stressful situations.
\end{itemize}

\subsection*{Likert Scale}
All questions use a 5-point Likert scale:
\begin{itemize}
    \item \textbf{1 = Strongly Disagree}
    \item \textbf{2 = Disagree}
    \item \textbf{3 = Neutral}
    \item \textbf{4 = Agree}
    \item \textbf{5 = Strongly Agree}
\end{itemize}

\subsection*{Questionnaire Items}
\begin{enumerate}
    \item I believe that most things can be accomplished if I put in the effort. (I)
    \item Luck plays a significant role in determining my success or failure. (E)
    \item When faced with difficulties, I actively seek ways to solve problems. (I)
    \item Even if I try very hard, the outcome of things is usually not within my control. (E)
    \item My efforts determine the outcome of things more than external circumstances. (I)
    \item When things go wrong, it's usually due to external circumstances or the influence of others. (E)
    \item When I feel stressed, I calm myself down through deep breathing.
    \item When I feel stressed, I want to know my own stress level.
    \item I like to know all the steps clearly before starting a task.
    \item I feel uneasy without clear operational procedures.
    \item When encountering complex tasks, I wish to be told what to do next.
    \item Compared to free exploration, I prefer to follow a strict task sequence.
    \item In high-pressure situations, I prefer clear external guidance.
    \item When I am under a lot of stress, I wish someone would comfort me.
    \item If someone encourages me, I have more confidence to complete the task.
    \item When I am emotionally unstable, I wish someone would help me calm down.
    \item When under great stress, I like to have someone by my side, even if they don't speak.
    \item When I feel anxious, I prefer a quieter and more controllable environment.
    \item Changes in the sound environment can significantly affect my level of concentration.
\end{enumerate}

\subsection*{Methodology for Interpretation}
\begin{itemize}
    \item \textbf{Questions 1-6 (LoC):} If the score for items marked `E' is higher than for items marked `I', the `Enhanced Self-Regulation' direction is not activated.
    \item \textbf{Questions 7-8 (IERQ - specific):} These appear only after the 'Enhanced Self-Regulation' direction is activated. If the score is below 3, the corresponding strategy (breathing guidance or stress feedback) is not activated.
    \item \textbf{Questions 9-13 (PNS):} All items are positively scored. If the total score is less than 15, the `Operational Process Guidance' direction is not activated. Time thresholds are 10/15/30 seconds:
        \begin{itemize}
            \item Total Score $\geq 22 \rightarrow 10$s
            \item Total Score $18 - 21 \rightarrow 15$s
            \item Total Score $<18 \rightarrow 30$s
        \end{itemize}
    \item \textbf{Questions 14-17 (IERQ - specific):} Determine whether the `NPC' intervention is activated.
    \item \textbf{Questions 18-19 (IERQ - specific):} Determine whether `Noise Reduction' is activated.
\end{itemize}

\section{Demographic information of the User Study participants.}\label{appendix:D}

\begin{table}[h]
    \centering
    \small
    \begin{tabular}{@{}lllll@{}}
        \toprule
        ID & Training Level & Exp. & Department & Gen. \\ \midrule
        P1  & Resident    & <1 year   & General Surgery      & F \\
        P2  & Resident    & 1-3 years & ENT / Otolaryngology & F \\
        P3  & Resident    & 1-3 years & General Surgery      & F \\
        P4  & Med Student & <1 year   & Clinical Rotation    & F \\
        P5  & Resident    & 1-3 years & ENT / Otolaryngology & F \\
        P6  & Resident    & <1 year   & Anesthesiology       & M \\
        P7  & Med Student & <1 year   & Clinical Rotation    & M \\
        P8  & Resident    & 1-3 years & Intensive Care Unit  & M \\
        P9  & Resident    & 1-3 years & Emergency Medicine   & F \\
        P10 & Resident    & <1 year   & ENT / Otolaryngology & M \\
        P11 & Resident    & <1 year   & Anesthesiology       & M \\
        P12 & Med Student & <1 year   & Clinical Rotation    & M \\
        P13 & Resident    & 1-3 years & General Surgery      & F \\ \midrule
        P14 & Resident    & 1-3 years & Neurology            & F \\
        P15 & Resident    & 1-3 years & Intensive Care Unit  & M \\
        P16 & Resident    & <1 year   & ENT / Otolaryngology & F \\
        P17 & Med Student & <1 year   & Clinical Rotation    & M \\
        P18 & Resident    & 1-3 years & General Surgery      & M \\
        P19 & Resident    & 1-3 years & Intensive Care Unit  & M \\
        P20 & Med Student & <1 year   & Clinical Rotation    & F \\
        P21 & Resident    & <1 year   & General Surgery      & M \\
        P22 & Resident    & <1 year   & Anesthesiology       & F \\
        P23 & Resident    & <1 year   & Emergency Medicine   & F \\
        P24 & Resident    & 1-3 years & Anesthesiology       & M \\
        P25 & Resident    & 1-3 years & General Surgery      & F \\
        P26 & Resident    & <1 year   & ENT / Otolaryngology & F \\ \bottomrule
    \end{tabular}
\end{table}

\newpage
\section{Result of Post-test Questionnaires}\label{appendix:E}
\begin{figure}[htbp]
    \centering
    \includegraphics[width=1\linewidth]{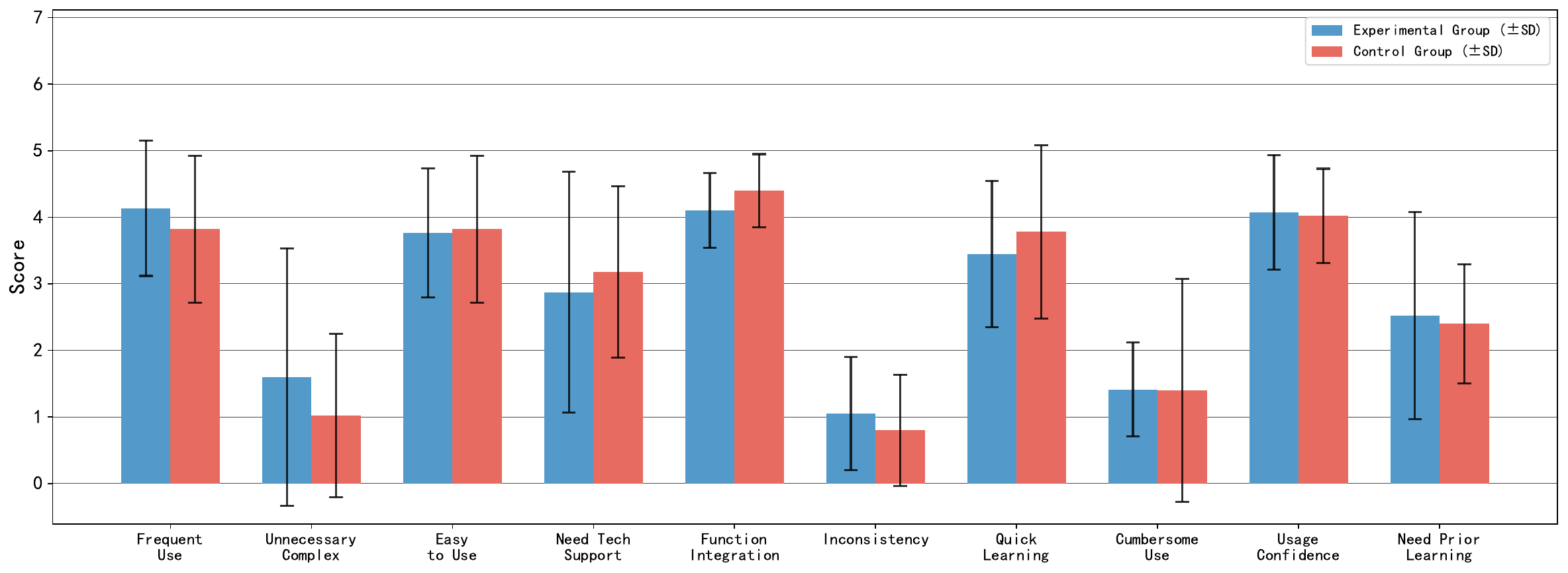}
    \caption{Results for the SUS portion of the post-simulation questionnaire. }
    \Description{A bar chart displaying the results of the System Usability Scale (SUS) questionnaire, comparing an Experimental Group (blue) and a Control Group (red). The vertical axis represents scores from 0 to 5, while the horizontal axis lists ten usability criteria, including 'Frequent Use,' 'Easy to Use,' and 'Function Integration.' For each criterion, two bars show the mean scores for each group, with error bars indicating the standard deviation. The scores between the two groups are very similar across all criteria, illustrating the text's finding that there was no significant difference in system usability between the groups.}
    \label{fig:sus}
\end{figure}

\begin{figure}[htbp]
    \centering
    \includegraphics[width=1\linewidth]{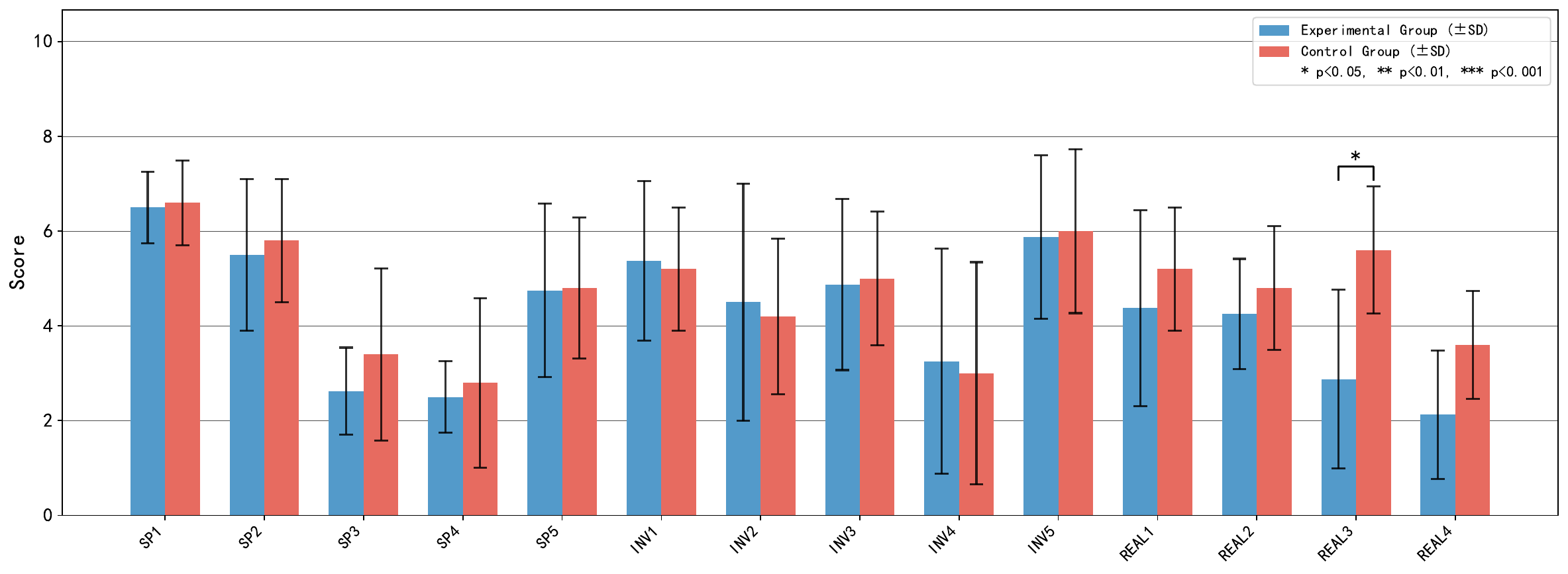}
    \caption{Results for the IPQ portion of the post-simulation questionnaire.}
    \Description{A bar chart displaying the results for the Igroup Presence Questionnaire (IPQ), comparing the mean scores of an experimental group (blue) and a control group (red-orange). The y-axis shows scores from 0 to 8, and the x-axis lists 14 different dimensions of presence and immersion, such as "Immersed," "Being There," and "VR World Reality." Error bars indicate the standard deviation for each group's score. Overall, the scores between the two groups are similar across most dimensions. However, the control group shows a noticeably higher mean score on the item "Hard to Distinguish from Reality.}
    \label{fig:ipq}
\end{figure}

\section{The Decision Logic of system}\label{appendix:F}

\begin{figure}[h]
    \centering
    \includegraphics[width=1\linewidth]{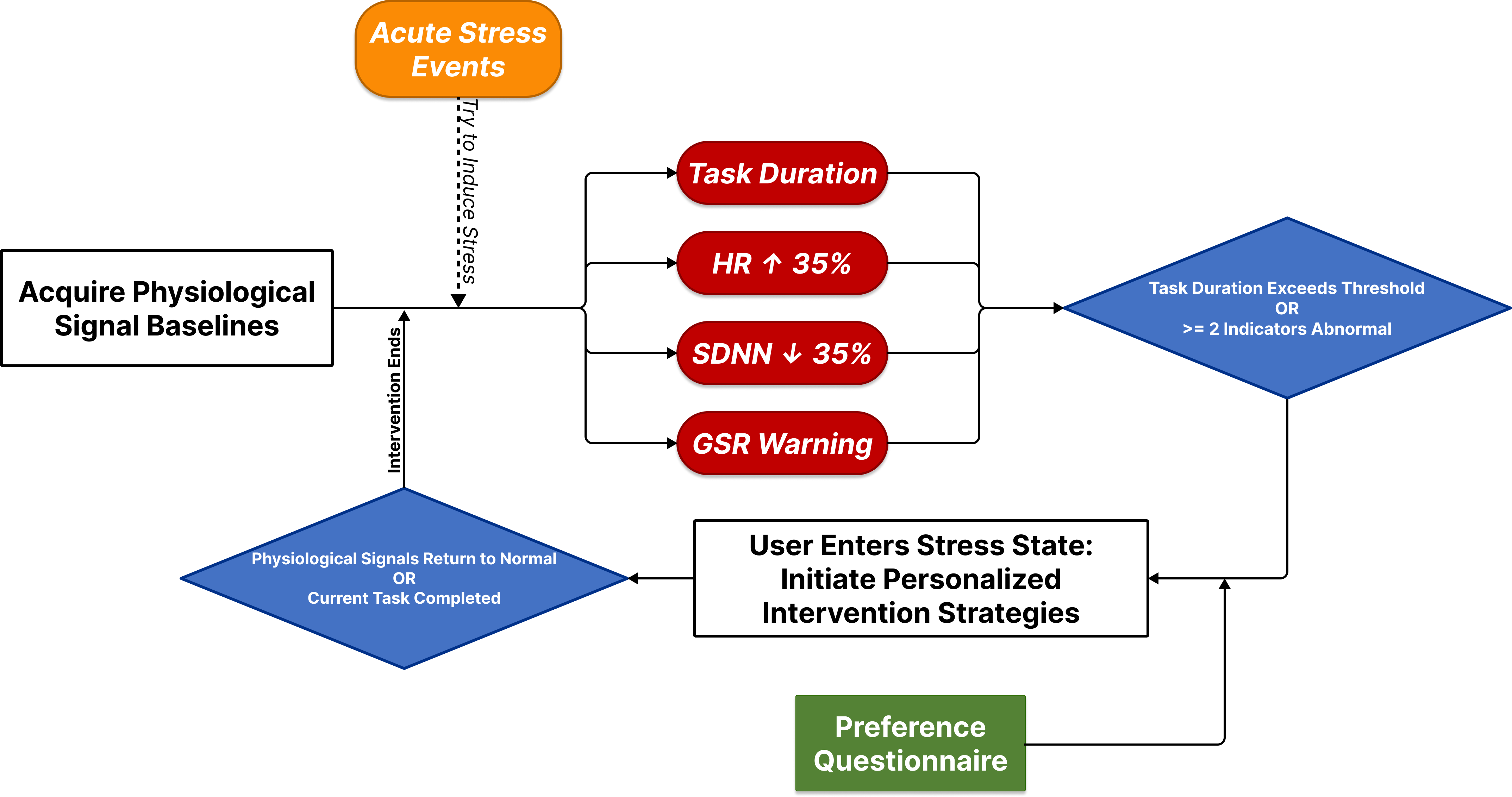}
    \caption{The decision-making workflow for stress detection and intervention triggering.}
    \Description{A flowchart illustrating the JITAI logic we used in our system. The process begins with 'Acquire Physiological Signal Baselines'. 'Acute Stress Events' attempt to induce stress. The system monitors four indicators: 'Task Duration', 'Heart Rate (HR) increase > 35\%', 'SDNN decrease > 35\%', and 'GSR Warning'. A decision diamond evaluates if 'Task Duration Exceeds Threshold OR >= 2 Indicators Abnormal'. If Yes, the system moves to 'User Enters Stress State', initiating personalized intervention strategies based on a 'Preference Questionnaire'. The intervention loops until a check confirms 'Physiological Signals Return to Normal OR Current Task Completed', at which point the intervention ends and the system returns to baseline monitoring.}
    \label{fig:strategy}
\end{figure}
\end{document}